\documentclass[11pt]{article}   
\pdfoutput=1

\usepackage{amsthm,amsmath,amssymb,bbold,xcolor,sectsty,latexsym,mathtools,hyperref,mathrsfs,slashed,bm,epsfig,enumerate}

\usepackage[textwidth=6.5in,top=1.2in,bottom=1.2in]{geometry}

\newcommand{\beq}{\begin{equation}}
\newcommand{\eeq}{\end{equation}}
\newcommand{\bea}{\begin{eqnarray}}
\newcommand{\eea}{\end{eqnarray}}
\newcommand{\amp}{&\!\!\!}

\newcommand{\rr}{R}
\newcommand{\R}{\tilde{R}}
\newcommand{\n}{\tilde{N}}
\newcommand{\Rm}{\tilde{R}_{\star}}
\newcommand{\E}{\tilde{H}}
\newcommand{\phiamp}{\phi_0}

\newcommand{\lamr}{\lambda_r}
\newcommand{\dphi}{\delta\phi}

\begin{document}

\thispagestyle{empty}
\begin{titlepage}
\nopagebreak

\title{ \begin{center}\bf Analysis of Dark Matter Axion Clumps\\ with Spherical Symmetry \end{center} }

\vfill
\author{Enrico D.~Schiappacasse$^{}$\footnote{enrico.schiappacasse@tufts.edu}, \,\,\, Mark P.~Hertzberg$^{}$\footnote{mark.hertzberg@tufts.edu}}

\date{\today}

\maketitle

\begin{center}
	\vspace{-0.7cm}
	{\it  $^{}$Institute of Cosmology, Department of Physics and Astronomy}\\
	{\it  Tufts University, Medford, MA 02155, USA}
	
\end{center}

\bigskip

\begin{abstract}

Recently there has been much interest in the spatial distribution of light scalar dark matter, especially axions, throughout the universe. When the local gravitational interactions between the scalar modes are sufficiently rapid, it can cause the field to re-organize into a BEC of gravitationally bound clumps. While these clumps are stable when only gravitation is included, the picture is complicated by the presence of the axion's attractive self-interactions, which can potentially cause the clumps to collapse. Here we perform a detailed stability analysis to determine under what conditions the clumps are stable. In this paper we focus on spherical configurations, leaving aspherical configurations for future work. We identify branches of clump solutions of the axion-gravity-self-interacting system and study their stability properties. We find that clumps that are (spatially) large are stable, while clumps that are (spatially) small are unstable and may collapse. Furthermore, there is a maximum number of particles that can be in a clump. We map out the full space of solutions, which includes quasi-stable axitons, and clarify how a recent claim in the literature of a new ultra-dense branch of stable solutions rests on an invalid use of the non-relativistic approximation. We also consider repulsive self-interactions that may arise from a generic scalar dark matter candidate, finding a single stable branch that extends to arbitrary particle number. 

\end{abstract}

\end{titlepage}

\setcounter{page}{2}

\tableofcontents

\newpage

\section{Introduction}

The nature of the bulk of the mass in the universe remains one of the biggest mysteries in modern physics and cosmology. A range of observations, including large scale structure, CMB, lyman $\alpha$ forrest, galactic rotation curves, are beautifully fit by the inclusion of cold dark matter. However, its particle physics origin is presently unknown. Among the favored candidates has traditionally been the WIMP and the axion, both of which have excellent motivations from considerations of shortcomings in the Standard Model of particle physics; the heirarchy problem and unification of couplings has often motivated certain kinds of WIMPs, while the strong CP problem and unification with gravity within string theory has often motivated the axion. A decades long search for WIMPs in direct detection experiments and colliders in the most obvious regime of parameter space has so far been unsuccessful (although interesting parameter space remains available). While the most highly motivated regime of the QCD axion's parameter space has yet to be fully probed experimentally, though a range of interesting experiments are both underway and planned. 

In this paper we shall focus on the QCD axion, as well as axion-like-particles or generic light scalar dark matter, and examine their astrophysical properties. The axion is a field that acquires a mass in the early universe, after the QCD phase transition, and can then begin to act as a form of cold dark matter \cite{Preskill:1982cy,Abbott:1982af,Dine:1982ah,Kim:2008hd}. Its initial momentum distribution is not predicted uniquely as it is sensitive to the details of inflation (e.g., see Ref.~\cite{Hertzberg:2008wr}). In particular, if the axion is present during inflation, then the field is driven to be highly homogeneous on large scales. On the other hand, if the axion is not present during inflation, then the field remains inhomogeneous from one Hubble patch to the next as suggested by causality. In the latter case, the large fluctuations already present in the axion field after the QCD phase transition can allow the field to exhibit strong mode-mode interactions from gravitation and re-organize into a type of Bose-Einstein condensate (BEC), as orginally discussed in Refs.~\cite{Sikivie:2009qn,Erken:2011dz}. In the former case, the fluctuations are initially much smaller, allowing for growth via perturbation theory in the matter era, and, while it is less clear, it could conceivably form a type of BEC too in the late universe.

The form of this axion BEC is not a conventional BEC with long range order, but instead has only short range order as it is a type of {\em localized clump}, as explained in Ref.~\cite{Guth:2014hsa}. The reason for this is that the ground state is a bound state due to the attractive nature of gravitation which is driving the BEC formation. Such a bound state is ordinarily well described by the weak field Newtonian approximation, with a $V=-Gm^2/r$ gravitational potential (though for extreme parameters, one can find a strong gravity regime, see Ref.~\cite{Helfer:2016ljl}). Although solving for such a multi-particle system in quantum mechanics is usually quite difficult, in the high occupancy BEC regime, it is well described by {\em classical field theory}, which accurately captures the properties of the underlying quantum theory when expectation values are computed appropriately, as shown in Ref.~\cite{Hertzberg:2016tal}. In the classical field theory and in the non-relativistic limit, one is still solving a non-linear PDE, but approximations are available and convincingly show that the ground state is {\em stable} against collapse under gravity for any number of axions. This is similar to the fact that the ground state of the hydrogen atom is well behaved which is also controlled by a $\sim -1/r$ potential. These gravitationally bound clumps are also known in the literature by various names, including ``Bose stars" \cite{Kolb:1993zz} and (especially when gravity is treated relativistically) ``oscillatons" \cite{UrenaLopez:2001tw}, and can organize into ``miniclusters".

A significant complication and potential instability arises from the inclusion of the axion's self-interactions in the form of a cosine potential. If expanded for small field values, the leading interaction is an attractive quartic term $-\lambda\,\phi^4$. Such a term can potentially cause the axion clump to collapse. In fact in the absence of gravity, such a term is known to give rise to a collapse instability. In the particle language, it is connected to an attractive delta function interaction between pairs of axions $V\propto -\lambda\,\delta^3({\bf x})$. It is known that the ground state energy is well behaved in 1-dimension (giving rise to a soliton solution), but is unbounded from below in higher dimensions (this dimensional dependence is studied in Ref.~\cite{Desjacques:2017fmf}). We will also consider repulsive self-interactions $+\lambda\,\phi^4$, which may arise from a generic scalar particle with a renormalizable potential.

So when {\em both} gravity and self-interactions are included, the situation is less clear and will be addressed in this work (other work includes Refs.~\cite{Chavanis:2011zi,Chavanis:2011zm}). By using a combination of analytical and numerical methods to study the ground state, we find that for sufficiently (spatially) large clumps, gravity dominates, and the system is stable. While for sufficiently (spatially) small clumps, self-interaction dominates, and the system is unstable. And we identify the boundary between these two phases. We find a maximum number of axions can be be present in the clumps. We also consider repulsive self-interactions, which may be relevant to a generic scalar dark matter candidate, finding that the maximum number of particle constraint is relaxed.

Moreover, we address an intriguing claim in the literature that an entirely new branch of axion solutions exist involving very dense clumps \cite{Braaten:2015eeu,Eby:2016cnq}. We find that the non-relativistic approximation used in these paper's treatment of this branch is unjustified. Instead there do exist quasi-stable relativistic solutions that are highly dense and governed by the cosine potential (rather than gravity), known as ``axitons" \cite{Kolb:1993hw}, which are connected to a certain limit of the physical solutions we analyze; we clarify their place in phase space.

In a forthcoming paper \cite{Forthcoming} we will extend the above results to include possible resonant decay into photons, as well as to move beyond the ground state (true BEC) to higher eigenstates, described by some non-zero angular momentum.

The outline of this paper is as follows: 
In Section \ref{AxionFieldTheory} we describe the basics of axion field theory and take the non-relativistic limit. 
In Section \ref{GroundState} we search for and describe spherically symmetric clump solutions. 
In Section \ref{TimeEvolution} we numerically solve for the full time evolution of the clumps.
In Section \ref{Physical} we compute realistic parameters of clumps for the QCD-axion.
In Section \ref{relativistic} we examine the possibility of a very dense branch of solutions.
In Section \ref{Repulsive} we consider the case of repulsive self-interactions for non-axion scalar dark matter.
In Section \ref{Conclusions} we present our summary and outlook. 
Finally, in Appendix \ref{AppendixA} we compute the field's instability about a homogeneous background.

\section{Axion Field Theory}\label{AxionFieldTheory}

\subsection{Axion Basics}

The axion is a pseudo-Goldstone boson associated with a spontaneously broken PQ symmetry $U(1)_{\text{PQ}}$ introduced as a solution to the strong CP problem \cite{Peccei:1977hh,Weinberg:1977ma,Wilczek:1977pj}. Axions are described in field theory by a real scalar field $\phi(x)$ with the following relativistic Lagrangian density 
\beq
\mathcal{L} = \sqrt{-g}\left[\frac{1}{2} g^{\mu\nu}\nabla_{\mu}\phi\, \nabla_{\nu}\phi - V(\phi)\right]\,,
\label{axionlagrangiandensity}
\eeq
with potential
\beq
V(\phi)=\Lambda^4[1-\cos(\phi/f_a)]\,,
\eeq
where $f_a$ is the PQ symmetry breaking scale. In the standard axion ``window" its value is $f_a\lesssim 10^{12}$\,GeV to avoid over-closure of the universe (though higher values of $f_a$ may be allowed depending on the details of inflation). The axion mass is identified from expanding the above potential, which gives $m=\Lambda^2/f_a$. The overall scale of the potential $\Lambda$ is of the order the QCD scale, and more precisely, it is given by
\beq
\Lambda^2 = {\sqrt{m_u m_d}\over m_u+m_d}\,f_\pi m_\pi \approx 0.06\,\mbox{GeV}^2\,.
\eeq
For definiteness we shall often take the axion mass to be $m=10^{-5}$\,eV, with $f_a = 6 \times 10^{11}\,\text{GeV}$, as representative values. The potential $V(\phi)$ comes from non-perturbative QCD effects, which break the initial $U(1)_{\text{PQ}}$ symmetry down to its discrete subgroup $Z(N_{DW})$~\cite{Sikivie:1982qv}. 

At the time when the axion mass is comparable with the Hubble time, the axion field begins to roll down to one of the $N_{DW}$ degenerate minima and domain walls are formed separating the different vacua. These domain walls then are attached to the axionic cosmic strings formed in the PQ phase transition~\cite{Vilenkin:1982ks}. We will focus on the scenario in which the PQ phase transition happens after inflation, where the initial fluctuations in the axion field are large from one Hubble patch to the next. Then to avoid the so-called ``axionic domain wall problem"~\cite{Sikivie:1982qv, Kim:1986ax}, we focus on models with $N_{DW} = 1$. For this case, the string-wall network vanishes quickly by fragmentation and decaying in axions~\cite{Barr:1986hs}. The large inhomogeneity in the axion field is ideal for strong mode-mode coupling to lead to BEC and clump formation.

The case in which the PQ phase transition happens before inflation implies an initially very homogeneous initial axion field. The homogenous configuration can lead to perturbations that undergo parametric resonance, which we address in the Appendix. So in this case the initial axion field could still evolve to clumps, though its efficiency is expected to be suppressed compared to the former case.

\subsection{Non-Relativistic Limit}

Expressing the cosine function as an infinite power series, $V(\phi)$ is
\beq
V(\phi ) = {1\over 2}m^2\phi^2-m^2 f_a^2 \sum_{n=2}^{\infty}{(-1)^n\over(2n)!} \left(\phi\over f_a\right)^{2n}\,,
\label{axionpotential}
\eeq
The non-relativistic field theory approximation for axions is often very well justified (though we shall later discuss regimes in which it is not). In the non-relativistic regime it is useful to express the real field $\phi$ in terms of a complex scalar field $\psi$ according to 
\beq
\phi({\bf{x}},t)=\frac{1}{\sqrt{2m}}\left[e^{-imt}\psi({\bf{x}},t)+e^{imt}\psi^{*}({\bf{x}},t)\right]\,,
\label{phi}
\eeq
with $\psi$ slowly varying. We then insert this expression into the axion Lagrangian density Eq.~(\ref{axionlagrangiandensity}). In the non-relativistic regime, all terms proportional to a power of $e^{-imt}$ or $e^{imt}$ can be safely dropped since they rapidly oscillate and approximately time average to zero. The $n^{\mbox{\tiny{th}}}$ term in the power series expansion of the potential is  
\beq
\left(\phi\over f_a\right)^{2n} = {(2n)!\over(n!)^2}\left(\frac{\psi^* \psi}{2mf_a^2}\right)^n\,\,(+\,\,\mbox{rapid oscillations})\,.
\label{binomial}
\eeq

Taking $|\dot{\psi}|/m \ll |\psi|$ in the kinetic term in Eq.~(\ref{axionlagrangiandensity}), dropping rapidly oscillating terms, re-summing all residual terms in the potential, and using the weak field Newtonian metric $g_{00}=1+2\phi_N(\psi^*,\psi)$, we obtain the following non-relativistic Lagrangian density for $\psi$
\beq
\mathcal{L}_{nr} = \frac{i}{2}\left(\dot{\psi}\psi^{*}-\psi\dot{\psi}^{*} \right)-\frac{1}{2m}\nabla \psi^*\! \cdot\! \nabla \psi -V_{nr}(\psi,\psi^*) - m\,\psi^*\psi\,\phi_N(\psi^*,\psi)\,,
\label{nonrelativisticlagrangiandensity}
\eeq
where the non-relativistic effective potential comes from considering only the leading non-linearity as follows
\beq
V_{nr}(\psi,\psi^*) = -{\psi^{*2}\psi^2\over 16\,f_a^2}\,,
\eeq
which is valid for small field amplitudes. This is required so that the typical frequency of oscillation is governed by $m$, plus small corrections, as is required in a non-relativistic treatment. In Section \ref{relativistic} we will return to the full potential to incorporate relativistic effects. Note that the $Z(N_{DW})$ symmetry present in the original Lagrangian density, Eq.~(\ref{axionlagrangiandensity}), is lost when the non-relativistic approximation is applied (we shall return to this issue). 

In phase space, $\psi$ and $\psi^*$ can be treated as independent fields, and are in fact canonically conjugate to each other with momenta $\pi=i\,\psi^*$. By performing a Legendre transformation, the total non-relativistic Hamiltonian is expressed by the sum of the following 3 terms
\beq
H_{nr} = H_{kin} + H_{int} + H_{grav}\,,
\label{hamiltoniandensityschematic}
\eeq
where 
\bea
H_{kin} \amp\equiv\amp {1\over2m}\int d^3x\, \nabla \psi^*\! \cdot\! \nabla \psi\,, \label{Hkin}\\
H_{int} \amp\equiv\amp \int d^3x \, V_{nr}(\psi,\psi^*) \,,\label{Hint}\\
H_{grav} \amp\equiv\amp -\frac{Gm^2}{2}\int d^3x  \int d^3x' \frac{\psi^*({\bf{x}})\psi^*({\bf{x}}')\psi({\bf{x}})\psi({\bf{x}}')}{|{\bf{x}}-{\bf{x}}'|}\,,
\label{Hgrav}
\eea
and $G$ is the gravitational constant. Here $H_{kin}$, $H_{int}$, $H_{grav}$, represent the kinetic energy, the self-interaction energy, and the gravitational energy, respectively. Note that we have dropped the overall rest mass energy $N\,m$ term, which is merely a constant in the non-relativistic theory.

The full equation of motion is
\beq
i\,\dot\psi=-{\nabla^2\psi\over2m}-Gm^2\,\psi\!\int d^3x'\frac{\psi^*({\bf{x}}')\psi({\bf{x}}')}{|{\bf{x}}-{\bf{x}}'|}+{\partial\over\partial\psi^*}V_{nr}(\psi,\psi^*)\,.
\label{fulleqm}
\eeq
Note that the final term is $\partial V_{nr}/\partial\psi^* = -\psi^*\psi^2/(8\,f_a^2)$.
  
Finally, we note that in the non-relativistic limit, the local number density of particles, $n({\bf x})$, and local mass density, $\rho({\bf x})$, are given by $n({\bf{x}})  = \psi^*({\bf{x}})\psi({\bf{x}})$ and $\rho({\bf x}) = m\,\psi^*({\bf x})\psi({\bf x})$.

\section{Ground State at Fixed Particle Number}\label{GroundState}

Now we proceed to analyze the axion system to determine how the three different terms in the Hamiltonian of Eq.~(\ref{hamiltoniandensityschematic}) combine to produce stable solutions such as axion clumps. (Perturbations around a homogenous background is left to the Appendix.) Since there is no known exact analytical solution for the ground state, we can proceed by using an approximate variational method to estimate the ground state of the system. We will later solve the system numerically.

\subsection{Spherical Symmetry}\label{SphericalSymmetry}

In this paper we will focus on spherically symmetric configurations. In a forthcoming paper \cite{Forthcoming} we will consider configurations that are not spherically symmetric. There we will show explicitly that such states have higher energy than the ground state, associated with additional energy from angular momentum (and related corrections). (This is similar to the well known case of the hydrogen atom: The hydrogen atom's ground state is spherically symmetric, while eigenstates that are not spherically symmetric, described by some spherical harmonic $Y_{lm}$ with $l>0$, have a higher energy.) Physically this makes sense: the theory respects rotational invariance and so it is very reasonable to suppose that the ground state is also spherically symmetric. Moreover, the theory is that of a scalar field (no vector field) and there is no mechanism here to spontaneously break rotational symmetry in the ground state. In this paper our primary focus is on describing the true ground state for a fixed number of axions, it is therefore guaranteed to be spherically symmetric.

We can write the ground state configuration as
\beq
\psi_g(r,t)= \Psi(r)\,e^{-i\,\mu\,t}\,,
\label{psispherical}
\eeq
where the shape is specified by the function $\Psi=\Psi(r)$ which is taken to only be a function of radius and can be taken to be real, and $\mu$ is the chemical potential. It is straightforward to insert this into the above Hamiltonian. For the kinetic and self-interaction terms, the angular integrals are trivial, giving
\bea
H_{kin} \amp = \amp {2\,\pi\over m}\int_0^\infty dr\,r^2\left(d\Psi\over dr\right)^2\,,\label{Hkinspherical}\\
H_{int} \amp = \amp 4\,\pi\int_0^\infty dr\,r^2\,V_{nr}(\Psi,\Psi)\,.\label{Hintspherical}
\eea
To compute $H_{grav}$ it is useful to use the spherical expansion for the inverse distance referred to a single origin of coordinates according to
\beq
\frac{1}{|{\bf x}-{\bf x}'|}= \sum_{l=0}^{\infty}\frac{4\pi}{2l+1}\left( \frac{r_<^l}{r_{>}^{l+1}} \right) \sum_{m=-l}^{l}Y_{l}^{{m}^*}(\theta,\varphi)Y_{l}^m(\theta',\varphi')\,,
\label{inversedistanceexpansion}
\eeq
where $r_<$ is the lesser and $r_>$ is the greater of $r=|{\bf x}|$ and $r'=|{\bf x}'|$. This shows that when we integrate over angles, only the $l=m=0$ terms survive. The gravitational contribution is then
\beq
H_{grav}=-{Gm^2\over2}(4\pi)^2\int_0^\infty dr\,r^2\int_0^\infty dr'\,r'^2{\Psi(r)^2\Psi(r')^2\over r_{>}}\,.
\label{Hgravspherical}
\eeq

\subsection{Simple Ansatz}\label{SimpleAnsatz}

Consider the time independent field equation for a spherically symmetric eigenstate. This takes the form
\beq
\mu\,\Psi=-{1\over 2m}\left(\Psi''+{2\over r}\Psi'\right)-4\pi Gm^2\Psi\int_0^\infty dr'\,r'^2\,{\Psi(r')^2\over r_{>}} + {1\over2}{\partial\over\partial\Psi}V_{nr}(\Psi)\,.
\label{timeindepschrodinger}\eeq
Let us begin by analyzing the far field region. For a bound state solution, the field must fall away rapidly at large radius, i.e., $\Psi\to0$ as $r\to\infty$. Hence at large distances we can ignore the self-interactions which are non-linear and behave as $\propto\Psi^3$ for small $\Psi$. Furthermore, in the gravitational term we can replace $r_{>}\to r$ in the far region, and then factorize for $1/r$, leaving an integral that gives the total number of particles in the clump $N =4\pi\int_0^\infty dr'\,r'^2\,\Psi(r')^2$. Hence
\beq
\mu\,\Psi \approx -{1\over 2m}\left(\Psi''+{2\over r}\Psi'\right)-{Gm^2N\over r}\Psi\,\,\,\,\,(\mbox{far region})\,.
\label{farregionschrodinger}\eeq
This is identical to the structure of the time independent Schr\"odinger equation for the hydrogen atom under replacement $Gm^2N\to e^2$. The spherically symmetric solutions eigen-modes are of the form
\beq
\Psi(r)=\mbox{Poly}_n(r)\times e^{-Gm^3N\,r/n}\,\,\,\,\,(\mbox{far region})\,.
\eeq
where $n=1,2,3,\ldots$ and $\mbox{Poly}_n(r)$ is a polynomial of degree $n$.

In the near field region, this obviously fails as the corrections from self-interactions become important and the structure of the gravitational term is altered. There are no known full analytical solutions. However, for the purposes of understanding qualitatively, and semi-quantitatively, the behavior of the system, it suffices to consider a simple ansatz for $\Psi$ throughout all space. A simple choice is to just use an exponential, with a decay length scale $\rr$ that is left free and acts as a variational parameter. We can write this as
\beq
\Psi_\rr(r)= \sqrt{\frac{N}{\pi\,\rr^3}}\,e^{-r/\rr}\,\,\,\,\,(\mbox{exponential ansatz})\,.
\label{psitrial}
\eeq
This ansatz has the disadvantage that it cannot be correct for small $r$. In particular, for small $r$ the field must have its derivative go to zero, i.e., $\Psi'\to0$ as $r\to0$ to ensure the first derivative term from the Laplacian $2\Psi'/r$ does not diverge. Hence there do exist more accurate solutions; we shall return to this in Section \ref{Improved}. We will find that the exponential ansatz is nonetheless correct to $\lesssim10\%$ in capturing the properties of the system, such as the ground state energy, etc, and will be used at various times in this paper.

The total number of particles, $N = \int d^3x\,n({\bf{x}})$, is ensured by the prefactor of Eq.~(\ref{psitrial}) and is assumed to be fixed as we perform our variation. Inserting the ansatz into Eqs.~(\ref{Hkinspherical},\,\ref{Hintspherical},\,\ref{Hgravspherical}), allows us to analytically obtain the value of the Hamiltonian within this ansatz. The kinetic $H_{kin}$ and gravitational $H_{grav}$ terms are readily evaluated, and so too is the self-interaction having replaced the cosine potential by its leading contributions (we shall return to the full potential in Section \ref{relativistic}). The total Hamiltonian is readily obtained as
\beq
H_{nr}(\rr)=\frac{N}{2m\rr^2} - \frac{5Gm^2N^2}{16\rr} -{N^2\over 128\pi f_a^2\,R^3}\,.
\label{hnonrel}
\eeq

It is useful to identify dimensionless quantities to simplify the analysis. We can define a dimensionless clump size $\R$, a dimensionless particle number $\n$ (we note that in classical field theory, without setting $\hbar=1$, $N=\int d^3x\,|\psi|^2$ actually has units of energy-time), and a dimensionless energy $\E$ as follows
\bea
\R \amp \equiv \amp mf_a \sqrt{G}\,\rr \,\,\,\,\,\, \mbox{(re-scaled clump size)}\,,\label{rescR}\\
\n \amp \equiv \amp \frac{m^2\sqrt{G}}{f_a}\,N \,\,\,\,\,\,\,\mbox{(re-scaled particle number)}\,,\label{rescN}\\
\E \amp \equiv \amp {m\over f_a^3\sqrt{G}}\,H_{nr}\,\,\,\,\,\,\,\,\, \mbox{(re-scaled energy)}\label{rescH}\,.
\eea
The dimensionless version of the Hamiltonian is then
\beq
\E(\R)=\frac{\n}{2\R^2} - \frac{5\n^2}{16\R} - \frac{\n^2}{128\pi\,\R^3 } \,.
\label{HamDimensionless}
\eeq

\subsection{Stable and Unstable Branches}

Extremizing the Hamiltonian $\E$ with respect to $\R$, we obtain the condition for stationary solutions
\beq
\R - \frac{5}{16}\n\R^2 - \frac{3}{128 \pi}\n= 0\,,
\label{zeroorderinY}
\eeq
whose solutions are simple
\beq
\R = \frac{8}{5\n} \pm \frac{\sqrt{512\pi-15\n^2}}{10\sqrt{2\pi}\,\n}\,.
\label{twosolutions}
\eeq

\begin{figure}[t]
\centering
\includegraphics[scale=0.36]{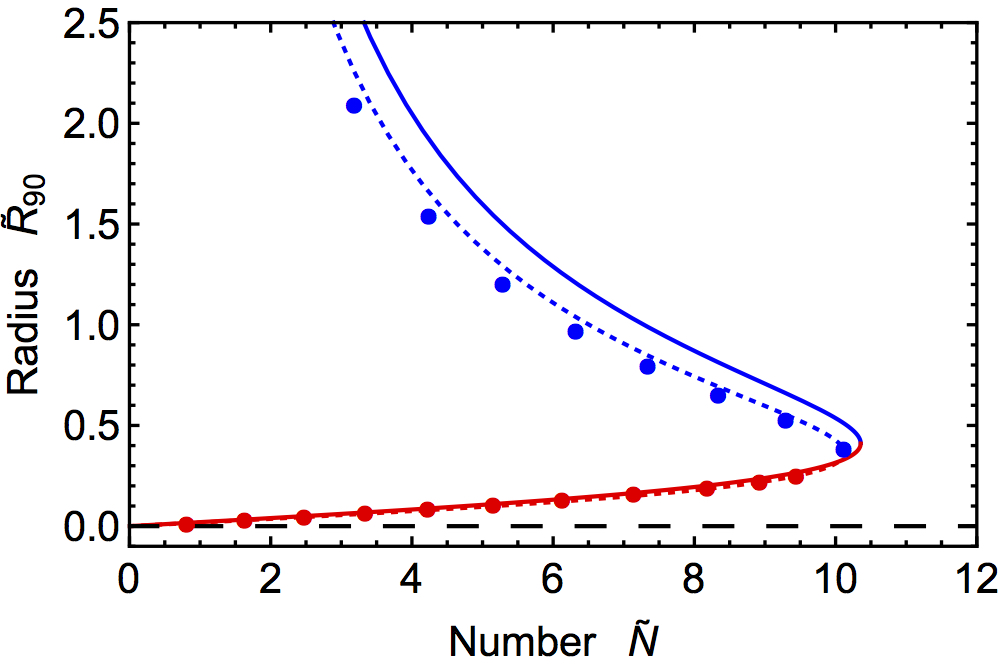}
\caption{Two branches of solutions when the system is treated in the non-relativistic regime for dimensionless radius $\R_{90}$ (defined as the radius that encloses 90\% of the mass) versus dimensionless particle number $\n$. The upper blue curve corresponds to stable solutions, while the lower red curve corresponds to unstable solutions. Solid curve is the exponential approximation, dotted curve is the sech approximation, and the individual dots are the exact numerical values. The condition for validity of the non-relativistic treatment is that one stays above the dashed curve (see ahead to Eq.~(\ref{rhomin})) this is valid for almost all points here, except in the very far left lower corner, which shall be treated in Section \ref{relativistic}, whose zoomed in behavior can be seen in Fig.~\ref{relativisticbranch}.}
\label{solutionsminimizedhamiltonian}
\end{figure}

In Fig.~\ref{solutionsminimizedhamiltonian} we plot these solutions. We have introduced a re-scaled value of $\R$, called $\R_{90}$, which is defined as the radius at which 90\% of the mass is enclosed, i.e.,
\beq
0.9\,N=4\pi\int_0^{\rr_{90}}dr'\,r'^2\,\Psi(r')^2\,.
\eeq
For the exponential ansatz, one finds $\rr_{90} \approx 2.661\,\rr$.

There are two branches of solutions: the stable one (blue line) is given by the dominance of gravity over the self-interaction and the unstable one (red line) is given by the opposite situation. This dominance is parametrically more pronounced as $\n$ decreases. By contrast, when $\n$ increases the gravitating and self-interacting terms become comparable. 

The solutions are restricted to the region 
\beq
\n < \n_{max} = \sqrt{512\pi\over15}\approx 10.36\,,
\label{nmaxexp}
\eeq
because for larger values of $\n$, the square root in Eq.~(\ref{twosolutions}) becomes imaginary. 

\begin{figure}[t]
\centering
\includegraphics[scale=0.28]{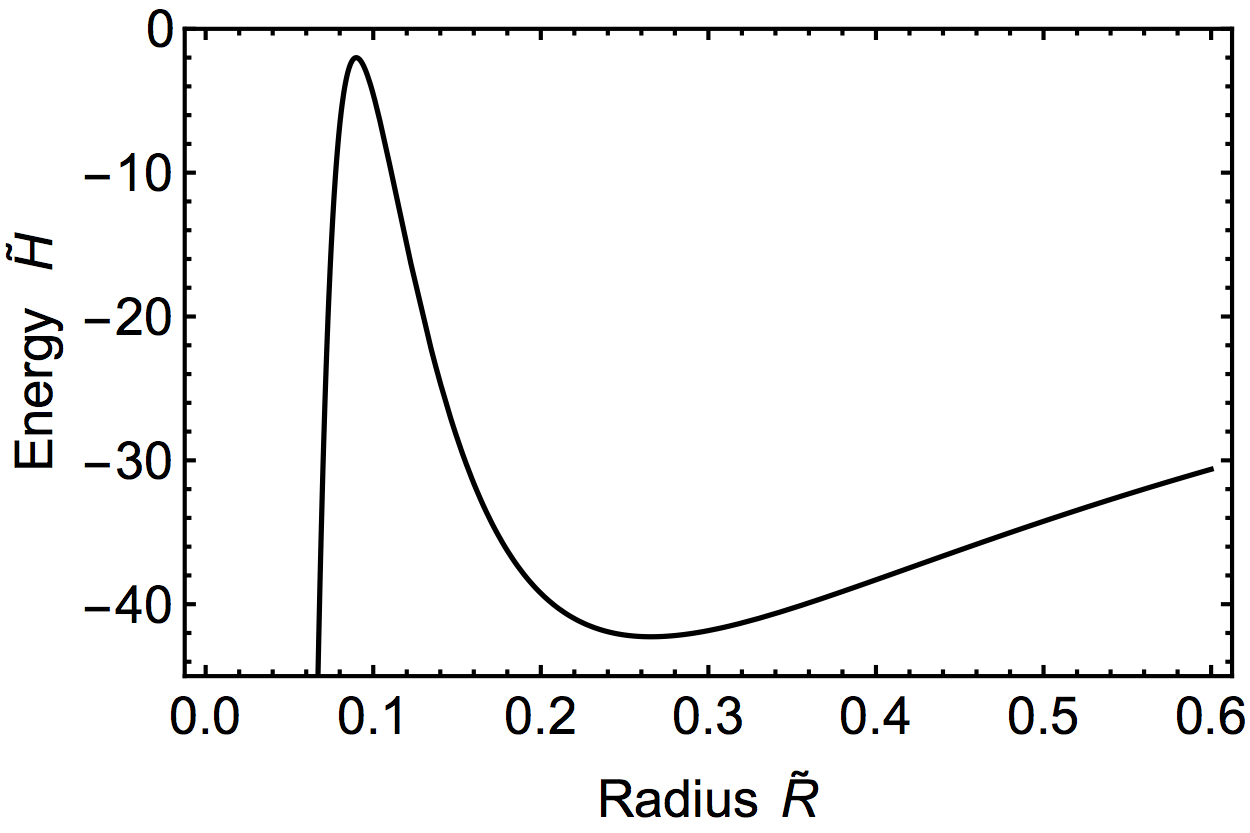}
\caption{A plot of the dimensionless energy $\E$ versus variational parameter (clump radius) $\R$ for a fixed value of particle number $\n=9$, within the exponential ansatz (other ansatzes look similar). The local maximum is associated with an unstable solution and corresponds to a point on the red curve of Fig.~\ref{solutionsminimizedhamiltonian}, while the local minimum is associated with a stable solution and corresponds to a point on the blue curve of Fig.~\ref{solutionsminimizedhamiltonian}.}
\label{hamiltonianvsradius}
\end{figure}

In Fig.~\ref{solutionsminimizedhamiltonian} we have labelled the upper (blue) curve as stable and the lower (red) curve as unstable. This is because the upper one corresponds to a local minimum of the Hamiltonian as a function of radius (at fixed particle number), while the lower one corresponds to a local maximum of the Hamiltonian. This is shown in Fig.~\ref{hamiltonianvsradius}, where we plot $\E=\E(\R)$ with $\n$ fixed at $\n=9$.

\subsection{Other Ansatzes}\label{Improved}

The above analysis used an exponential ansatz for the radial profile, which of course is not exact. Although an exponential fall off is correct at large $r$ (albeit with a modified fall-off rate), it is not true at small $r$. In the small $r$ regime, we can replace $r_>\to r'$ in Eq.~(\ref{timeindepschrodinger}), which changes the structure of this term to modify the effective chemical potential. For ease of presentation, let's ignore the self-interaction term temporarily, and then the equation becomes
\beq
\mu_{eff}\,\Psi\approx -{1\over 2m}\left(\Psi''+{2\over r}\Psi'\right)\,\,\,\,\,(\mbox{near region, ignoring $V_{nr}$})\,,
\label{nearregionschrodinger}
\eeq
where
\beq
\mu_{eff} = \mu+4\pi Gm^2\int_0^\infty dr'\,r'\,\Psi(r')^2\,.
\eeq
It is anticipated that this $\mu_{eff}$ is positive and then the solutions of Eq.~(\ref{nearregionschrodinger}) are spherical Bessel functions. The ground state is the Bessel function of order 0
\beq
\Psi(r)\propto j_0(\sqrt{2m\mu_{eff}}\,r)={\sin(\sqrt{2m\mu_{eff}}\,r)\over \sqrt{2m\mu_{eff}}\,r}\,\,\,\,\,(\mbox{near region, ignoring $V_{nr}$})\,,
\eeq
Including self-interactions, this shape is corrected, but the salient feature that survives is that the solution is an inverted parabola centered at $r=0$, plus higher order corrections
\beq
\Psi(r)=\Psi_0-{1\over 2}\Psi_2\,r^2+\ldots\,\,\,\,\,(\mbox{near region})\,,
\eeq
where the values of $\Psi_0$ and $\Psi_2$ are actually sensitive to the {\em full} shape of the potential and so are not easily obtained.

A better ansatz than the above exponential is one that carries both of these features: exponential decay at large $r$ and inverted parabola around $r=0$. A couple of neat examples that still only carry a single variational parameter $R$ and satisfies these properties are
\bea
\Psi_R(r) \amp = \amp \sqrt{3N\over\pi^3R^3}\,\mbox{sech}(r/R)\,\,\,\,\,(\mbox{sech ansatz})\,,\\
\Psi_R(r) \amp = \amp \sqrt{N\over7\pi R^3}\,(1+r/R)\,e^{-r/R}\,\,\,\,\,(\mbox{exponential $\times$ linear ansatz})\,.
\eea
Inserting this into the Hamiltonian Eqs.~(\ref{Hkinspherical},\,\ref{Hintspherical},\,\ref{Hgravspherical})  and using dimensionless variables, we obtain a modified version of Eq.~(\ref{HamDimensionless})
\beq
\E(\R) = a\frac{\n}{\R^2} - b\frac{\n^2}{\R} - c\frac{\n^2}{\R^3 }\,,
\label{hamiltonianquadraticgeneral}
\eeq
where
\bea
&& a={12+\pi^2\over6\pi^2},\,\,\, b={6(12\,\zeta(3)-\pi^2)\over\pi^4},\,\,\, c={\pi^2-6\over 8\pi^5}\,\,\,\,\,(\mbox{sech ansatz})\,,\\
&& a={3\over14},\,\,\, b={5373\over25088},\,\,\, c={437\over200704\pi}\,\,\,\,\,(\mbox{exponential $\times$ linear ansatz})\,,
\eea
for the sech and exponential $\times$ linear ansatzes, respectively. In fact {\em any} localized ansatz of a single variational parameter $\rr$ can be put into this general form, with only the values of the coefficients $(a,b,c)$ sensitive to the ansatz's details. For any values of $(a,b,c)$ there is still a stable branch for large $\R$ and an unstable branch for low $\R$, given by a generalization of Eq.~(\ref{twosolutions}) to
\beq
\R = {a\pm\sqrt{a^2-3bc\n^2}\over b\n}\,.
\label{Rabc}
\eeq
For the sech ansatz, this is given as the dotted blue and red curves in Fig.~\ref{solutionsminimizedhamiltonian} (the exponential $\times$ linear is not plotted, but is found to be very slightly more accurate than the sech). Note that for the sech function, the radius that encloses 90\% of the mass is $\rr_{90}\approx 2.799\,\rr$, while for the exponential $\times$ linear function, it is $\rr_{90}\approx 3.610\,\rr$. The Hamiltonian looks qualitatively similar to Fig.~\ref{hamiltonianvsradius}. By extremizing the Hamiltonian, the maximum value of $\n$ is
\beq
\n<\n_{max}={a\over\sqrt{3bc}}\,,
\label{Nabc}
\eeq
giving $\n_{max}\approx 10.12$ for the sech and $\n_{max}\approx 10.15$ for the exponential $\times$ linear, and so both are within $\sim 2\%$ of the result of the exponential ansatz of Eq.~(\ref{nmaxexp}). 

Furthermore, we find that the (binding) energy of the ground state is lowered in this sech ansatz; as expected as it improves the physical behavior for small $r$. The energy is in general a slightly complicated function of $\R$, however at the critical point, where the two branches meet, it is
\beq
\E_{crit}=-{a^2\sqrt{b}\over 9\sqrt{3}\,c^{3/2}}\,,
\eeq
which lowers the energy of the ground state by $\sim 2\%$ from the exponential ansatz. 

\begin{figure}[t]
\centering
\includegraphics[scale=0.35]{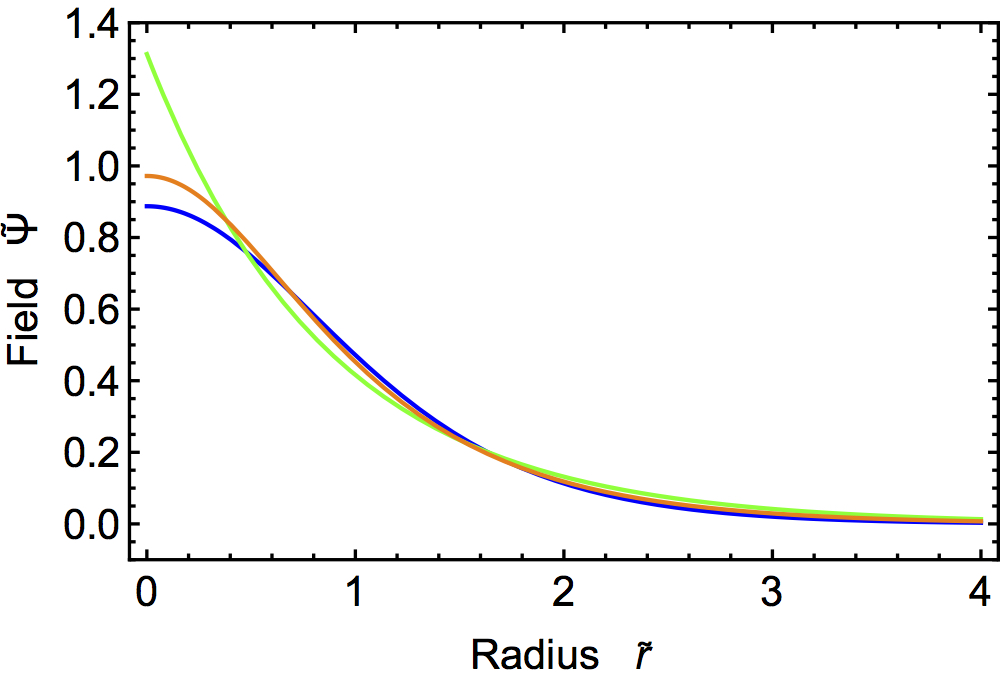}
\caption{Field $\tilde{\Psi}$ versus radius $\tilde{r}$ for the ground state at a fixed number of particles ($\n=3.565$) on the stable (blue) branch of Fig.~\ref{solutionsminimizedhamiltonian} in the non-relativistic theory. Blue is the exact numerical result, green is the exponential approximation, and orange is the sech approximation.}
\label{FieldvsRadius}
\end{figure}

For a specific value of $\n$ (namely $\n=3.565$), we plot the field $\Psi(r)$ in Fig.~\ref{FieldvsRadius} on the stable branch, with the exponential ansatz in green, the sech ansatz in orange, and we have also solved the equation of motion numerically to find the exact result in blue. We find that the sech tends to always be within a percent or so of the true energy, while the exponential can be a few percent worse. The exact numerical result for the phase diagram is indicated by the individual dots in Fig.~\ref{solutionsminimizedhamiltonian}. For the blue stable branch, the sech does considerably better than the exponential; while on the red unstable branch, the exponential does marginally better than the sech. In Section \ref{relativistic}, we will study the lower left corner of the phase diagram, and exploit the exponential ansatz to obtain some understanding of its behavior.

\section{Numerical Solution for Time Evolution}\label{TimeEvolution}

In this section we would like to compute the full nonlinear evolution of the axion field numerically (within the spherically symmetric phase space). We will demonstrate that by perturbing away from the above clump solutions, there is indeed a ``stable" branch and an ``unstable" branch, in agreement with the above descriptions.

\subsection{Numerical Recipe}
We would like to solve the full equation of motion for the axion field, Eq.~(\ref{fulleqm}), within the spherically symmetric ansatz. For gravitation it is useful to make use of the Poisson equation for the Newtonian potential. Working with dimensionless variables, this pair of equations is given by
\bea
&& i\,\frac{\partial\tilde{\psi}}{\partial\tilde{t}}=-\frac{1}{2\tilde{r}}\frac{\partial^2}{\partial \tilde{r}^2}\left( \tilde{r}\,\tilde{\psi} \right)   + {\tilde{\phi}}_N \tilde{\psi}- \frac{1}{8}|\tilde{\psi}|^2 \tilde{\psi}\,, \label{Aeq}\\
&&\frac{1}{\tilde{r}} \frac{\partial^2}{\partial \tilde{r}^2}\left( \tilde{r}\,\tilde{\phi}_N   \right)  =4\pi |\tilde{\psi}|^2\,,
\label{Peq}
\eea
where $\tilde{\psi}(\tilde{r},\tilde{t})$ and  $\tilde{\phi}_N(\tilde{r},\tilde{t})$ are the axion field and the Newtonian potential, respectively, and $\tilde{r}$  and $\tilde{t}$ are the radial and time coordinates, respectively, all in dimensionless variables. 

Lets define the spatial domain as $[\tilde{r}_{start},\tilde{r}_{end}]$, and the time domain as $[\tilde{t}_{initial},\tilde{t}_{final}]$. For numerical purposes, we discretize these domains as $\tilde{r}_l = \tilde{r}_{start} + l\, \Delta r$ for $l=0,\dots,n_r$, and $\tilde{t}_j =\tilde{t}_{initial}+ j\,\Delta t$ for $j=0,\dots,n_t$. Here $\Delta r$ and $\Delta t$ correspond to the radial and time step-sizes, respectively. The boundary conditions are taken to be $\partial \tilde{\psi}(\tilde{r}_{start},\tilde{t})/\partial \tilde{r}=\partial \tilde{\phi}_N (\tilde{r}_{start},\tilde{t})/\partial \tilde{r}=0$, with $\tilde{r}_{start}$ close to 0, and $\tilde{\psi}(\tilde{r}_{end},\tilde{t})=\tilde{\phi}_N(\tilde{r}_{end},\tilde{t})=0$, with $\tilde{r}_{end}$ much greater than the characteristic radius of the profile of the axion field to prevent unphysical reflection at the boundary. 

In order for the time evolution to be sufficiently stable, we use the Crank-Nicolson method (this method was used in Ref.~\cite{Harrison}, although no self-interactions were included in that work). In this method one splits the time derivative of the axion evolution equation by finite difference in the usual way, but specifies the right hand side as an average of the value at the $j$ time step and the $j+1$ time step, which schematically appears as follows
\beq
i\frac{\tilde{\psi}_l^{j+1}-\tilde{\psi}_l^j}{\Delta t}={1\over2}\left[F_l^{j+1}\left(\tilde{\phi}_N,\tilde{\psi},\tilde{r},\tilde{t},\frac{\partial \tilde{\psi}}{\partial\tilde{r}},\frac{\partial^2\tilde{\psi}}{\partial\tilde{r}^2}  \right) + F_l^{j}\left( \tilde{\phi}_N,\tilde{\psi},\tilde{r},\tilde{t},\frac{\partial \tilde{\psi}}{\partial\tilde{r}},\frac{\partial^2\tilde{\psi}}{\partial\tilde{r}^2} \right)   \right]\,.
\label{CN}
\eeq
where the $F$'s are implicitly defined by Eq.~(\ref{Aeq}). For all spatial derivatives, both in the axion $\psi$ and Newtonian potential $\phi_N$, we use a standard central difference method. Now the crucial point is that the right hand side has terms $F_l^{j+1}$, which includes terms proportional to $a_l^{j+1}\equiv\tilde{\phi}_{N,l}^{j+1}\tilde{\psi}_l^{j+1}$ and $b_l^{j+1}\equiv|\tilde{\psi}_l^{j+1}|^2\tilde{\psi}_l^{j+1}$, which are to be solved for. We apply an iterative method to find $\tilde{\psi}^{j+1}$ by solving alternately Eqs.~(\ref{Aeq},\,\ref{Peq}). We write $\tilde{\psi}_l^{j+1, q}$ and $\tilde{\phi}_{N,l}^{j+1, q}$ for the iterates, where $q$ is an index specifying the iteration step. We then use the following algorithm at each time step:
\begin{enumerate}[{(a)}]
\item Take $\tilde{\phi}_{N,l}^{j+1}=\tilde{\phi}_{N,l}^{j}$ in $a_l^{j+1}$ and $\tilde{\psi}_l^{j+1}=\tilde{\psi}_l^{j}$ in $b_l^{j+1}$
to obtain an updated value for the axion field, $\tilde{\psi}_l^{j+1,0}$, by solving Eq.~(\ref{CN}).
\item Use $\tilde{\psi}_l^{j+1,0}$ in Eq.~(\ref{Peq}) to obtain  $\tilde{\phi}_{N,l}^{j+1,1}$.
\item Use  $\tilde{\phi}_{N,l}^{j+1,1}$ in  $a_l^{j+1}$ and $\tilde{\psi}_l^{j+1,0}$ in $b_l^{j+1}$ to obtain $\tilde{\psi}_l^{j+1,1}$ by solving Eq.~(\ref{CN}).
\item Repeat steps (b) and (c) as many times as needed until the desired degree of convergence is reached by satisfying $|\tilde{\psi}_l^{j+1,q+1}-\tilde{\psi}_l^{j+1,q}| < T$, where $T$ is the desired tolerance.
\item If the desired tolerance is not reached at a defined number of maximum iterations, start from (a), but decrease the time step-size appropriately.
\end{enumerate}

\subsection{Stable and Unstable Branches}

\begin{figure}[t]
\centering
\includegraphics[scale=0.2]{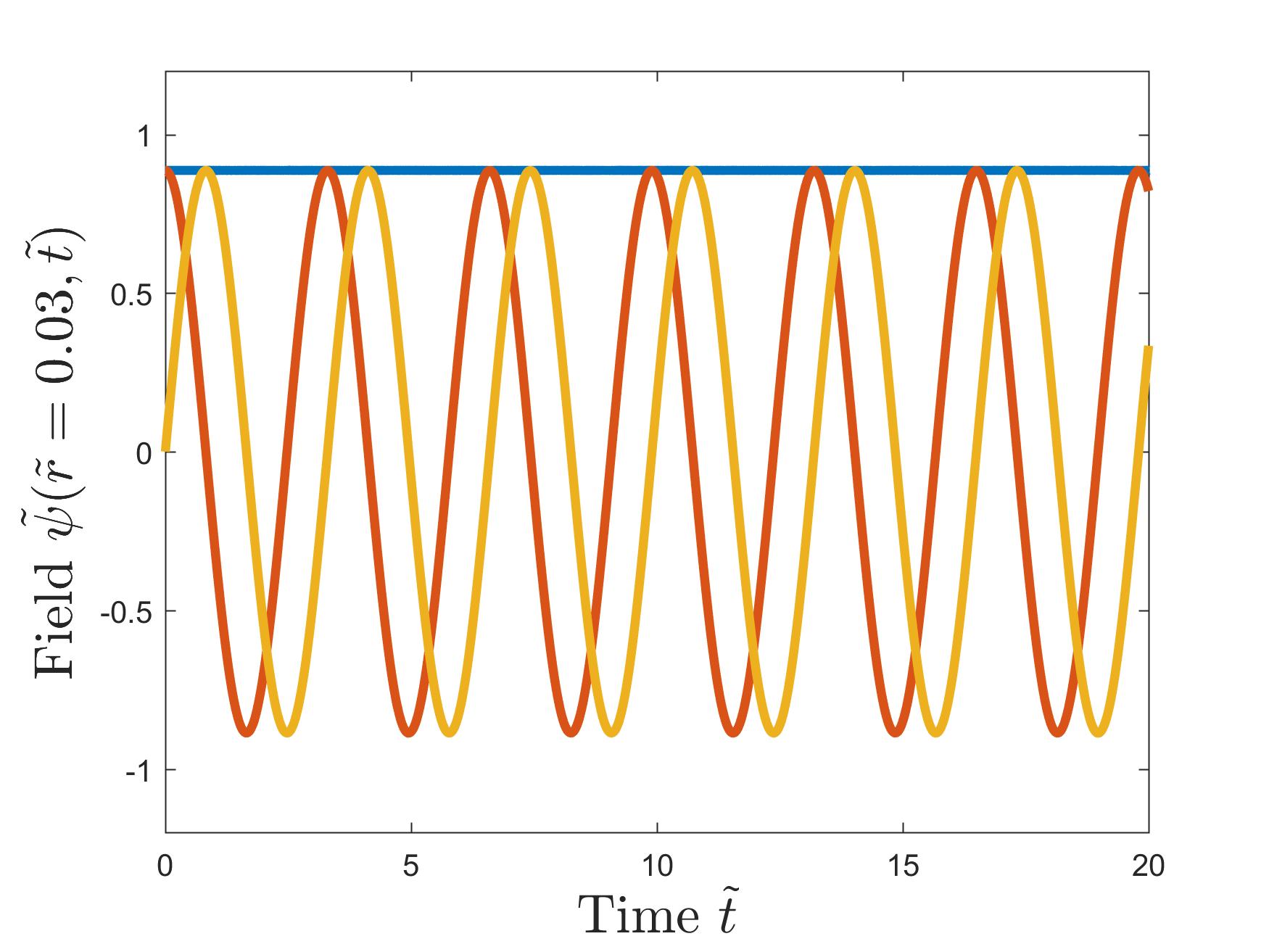}
\caption{Field $\tilde{\psi}$ at a fixed radius $\tilde{r}=0.03$ as a function of time $\tilde t$ for a given number of particles ($\n=3.565$) on the stable (blue) branch of Fig.~\ref{solutionsminimizedhamiltonian}. We have chosen initial conditions so that the field is real. Red is real part $Re[\tilde{\psi}]$, yellow is imaginary part of $Im[\tilde\psi]$, blue is the absolute value. We see that the field is periodic in time.}
\label{Evolve1stable}
\end{figure}

Using the above numerical recipe, we have solved for the time evolution of the axion system for both stable and unstable solutions.  In Fig.~\ref{Evolve1stable} we plot the time evolution of a clump that lives exactly on the stable branch solution. We show both the real and imaginary and absolute values of the (re-scaled) field $\tilde\psi$. Clearly the field is oscillating periodically, as a ground state solution should. Note that we only plot the stable branch here. In principle we can also plot a clump exactly sitting on the unstable branch, which is in principle is also periodic. However, any tiny numerical perturbations causes the solution to depart after some finite time, as we now discuss more systematically.

It is important to perturb these solutions by a finite amount and track its time evolution. As an explicit type of perturbation, we consider the following initial condition
\beq
\psi_{initial}(r)=(1+\epsilon)\Psi(r)\,,
\eeq
where $\Psi(r)$ is the (real) spherically symmetric clump solution. Here $\epsilon$ is taken to be a constant parameter that measures how far from the exact clump solution we begin. We are focussing here only on spherically symmetric perturbations for the following reasons: as we explained at the start of Section \ref{SphericalSymmetry}, the true ground state should be spherically symmetric and so will be stable against {\em all} perturbations. So we focus only on spherical perturbations, as aspherical perturbations provide additional energy from angular momentum, moving the state away from a global minimum in energy. Furthermore, the subject of aspherical axion configurations will be studied in our forthcoming paper \cite{Forthcoming}.

\begin{figure}[h!]
\centering
\includegraphics[scale=0.195]{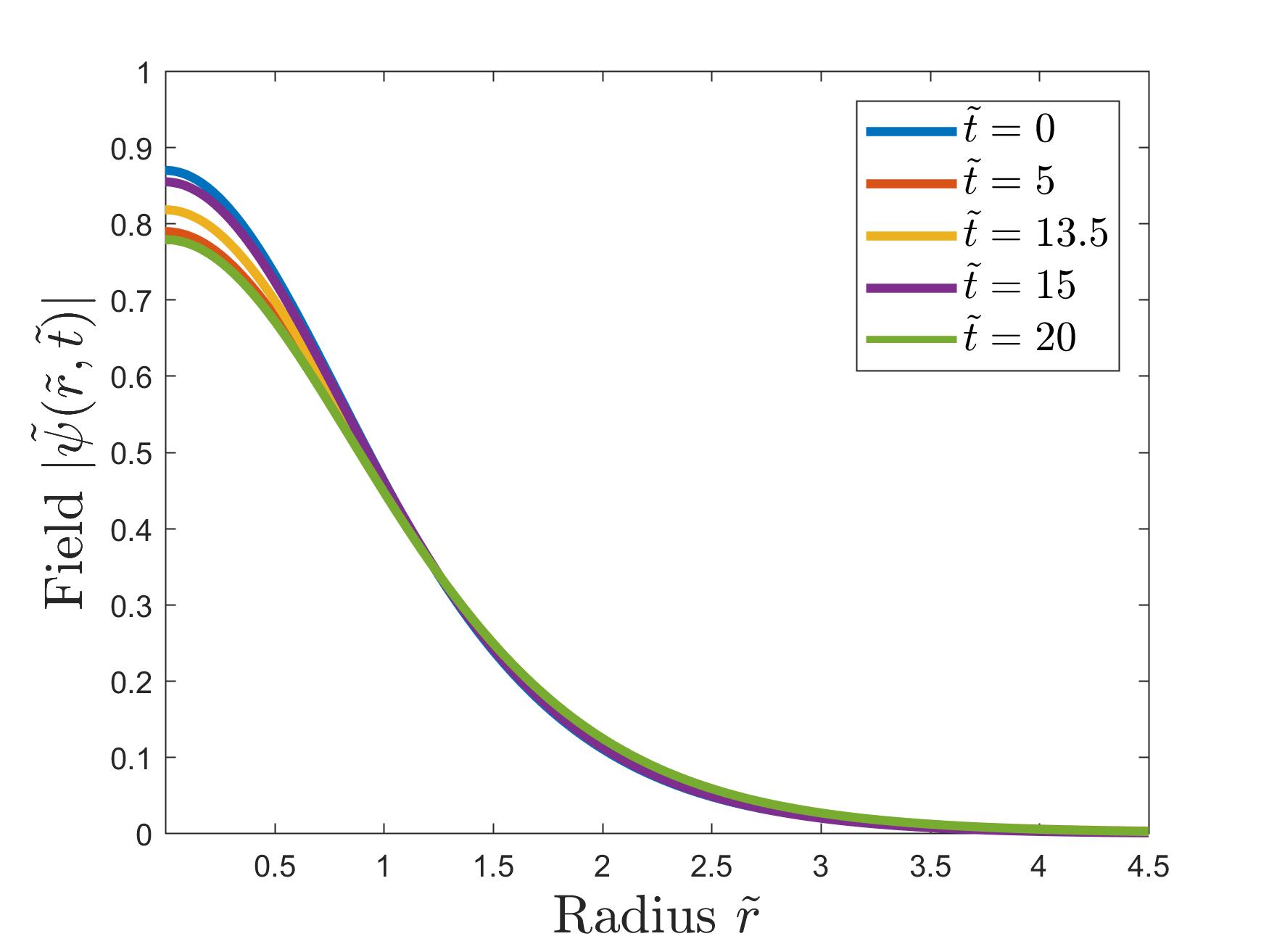}
\includegraphics[scale=0.195]{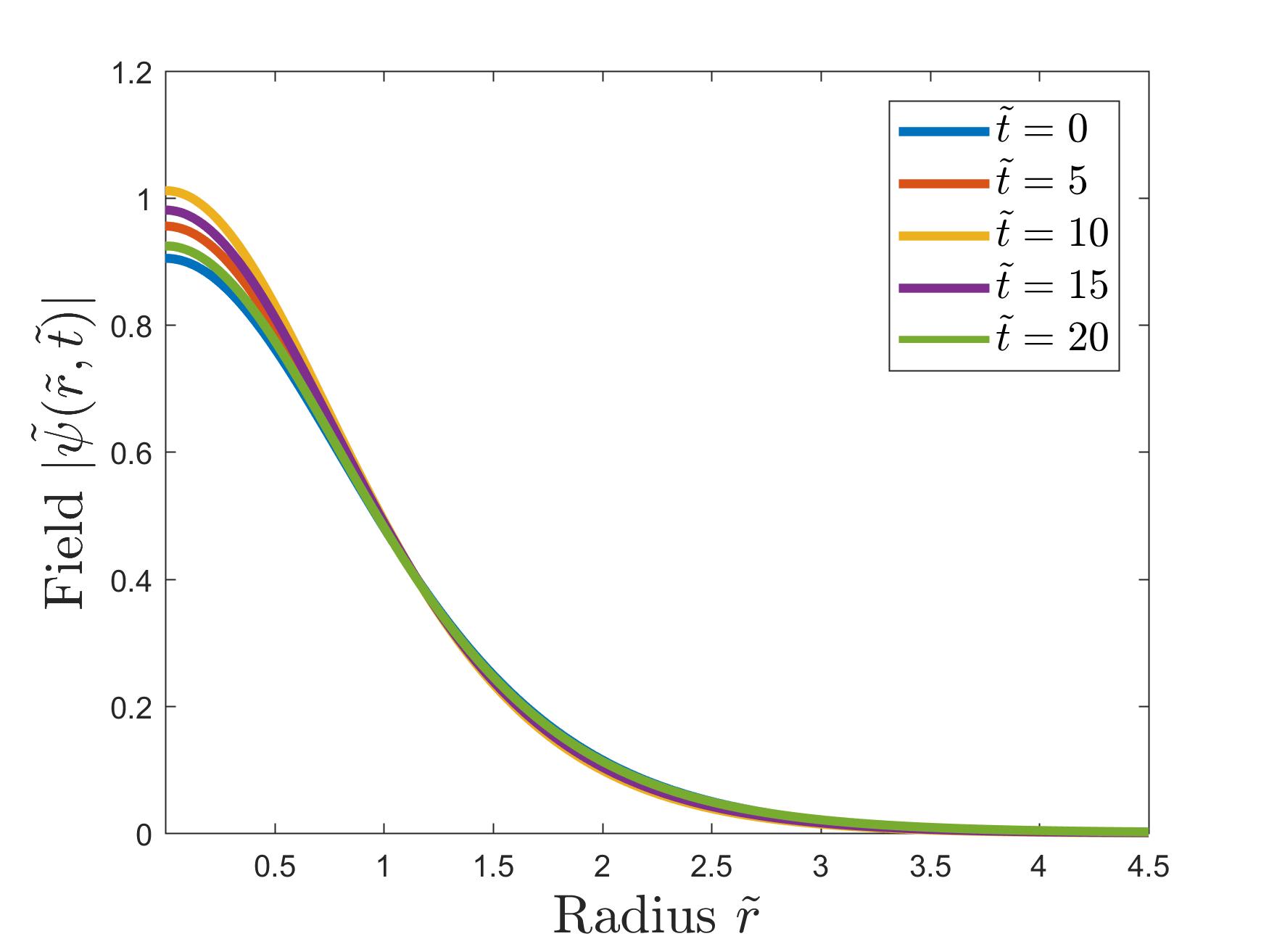}
\caption{Magnitude of field $|\tilde{\psi}|$ as a function of radius $\tilde{r}$ at different times $\tilde t$ for a given number of particles. Upper: we have perturbed $\tilde\psi$ away from the stable (blue) branch of Fig.~\ref{solutionsminimizedhamiltonian} with $\n=3.565$ by $-$2\%, giving $\n=3.424$. Lower: we have perturbed $\tilde\psi$ away from the stable (blue) branch of Fig.~\ref{solutionsminimizedhamiltonian} with $\n=3.565$ by +2\%, giving $\n=3.709$. We see that the field is indeed stable, since it merely oscillates in time.}
\label{Evolve2Stable2}
\end{figure}
\begin{figure}[h!]
\centering
\includegraphics[scale=0.195]{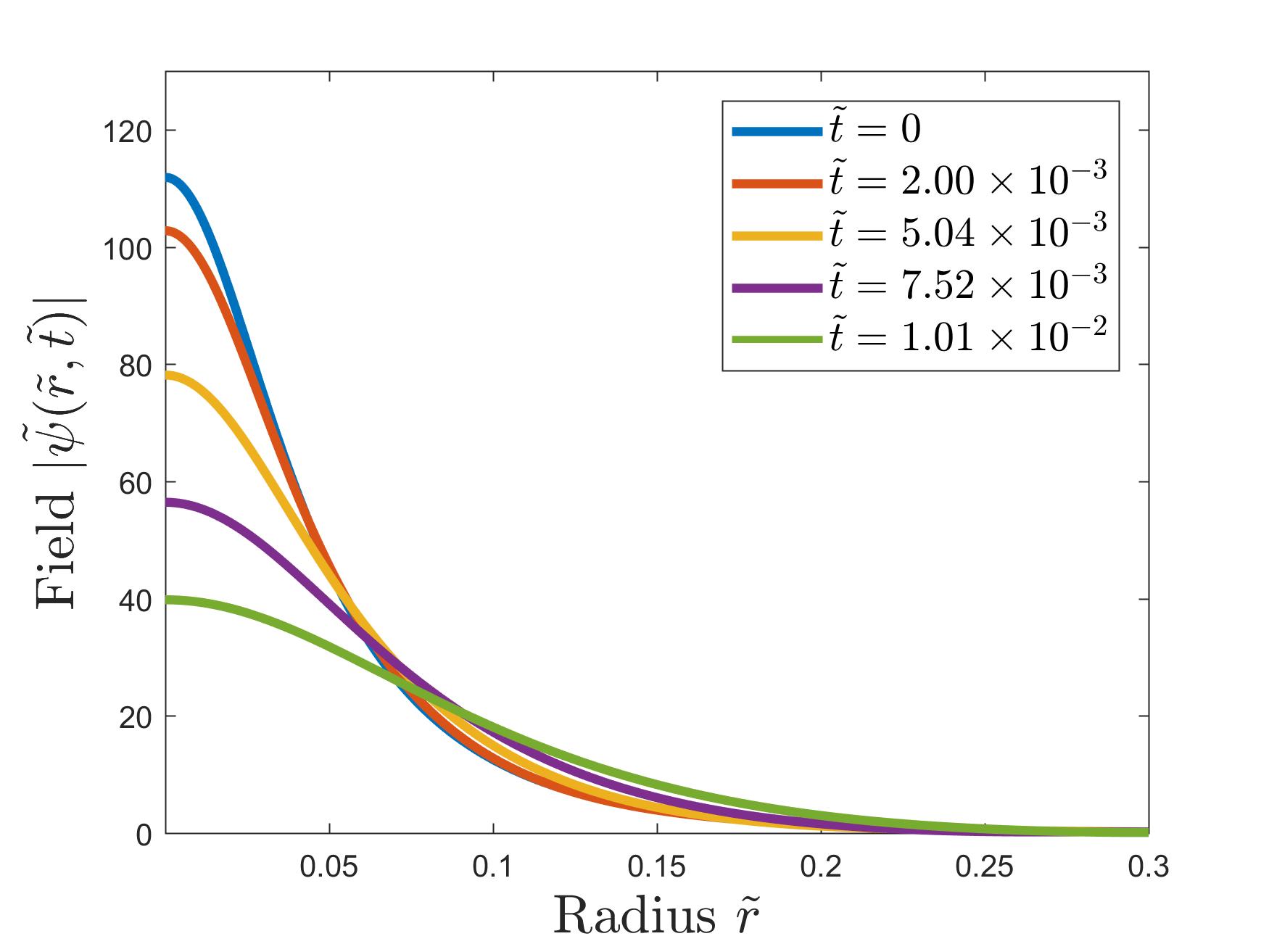}
\includegraphics[scale=0.195]{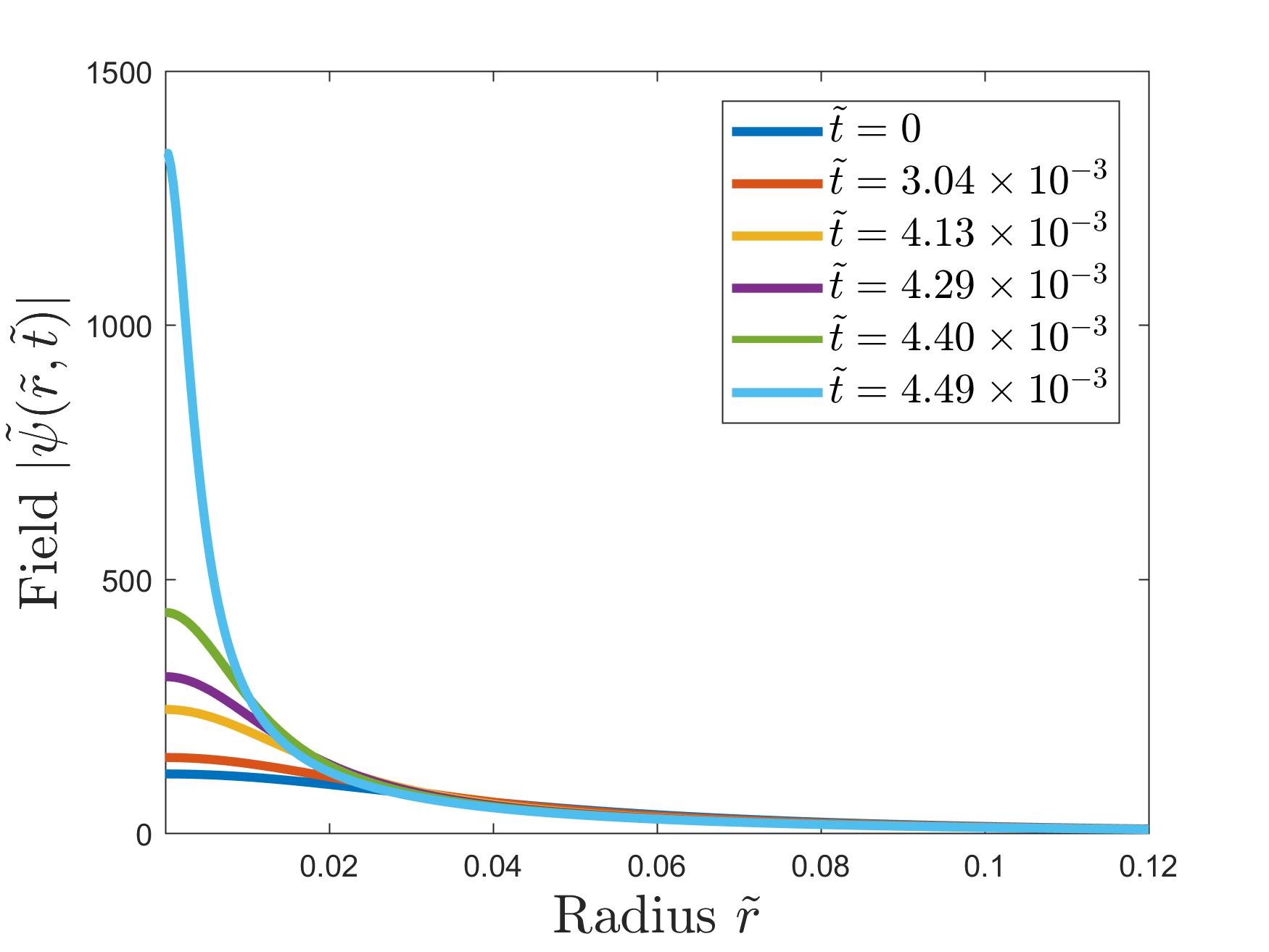}
\caption{Magnitude of field $|\tilde{\psi}|$ as a function of radius $\tilde{r}$ at different times $\tilde t$ for a given number of particles. Upper: we have perturbed $\tilde\psi$ away from the unstable (red) branch of Fig.~\ref{solutionsminimizedhamiltonian} with $\n=3.565$ by $-$2\%, giving $\n=3.424$. Lower: we have perturbed $\tilde\psi$ away from the unstable (red) branch of Fig.~\ref{solutionsminimizedhamiltonian} with $\n=3.565$ by +2\%, giving $\n=3.709$. We see that the field is indeed unstable, since it expands outwards in upper and collapses inwards in lower.}
\label{Evolve2Unstable2}
\end{figure}

In Fig.~\ref{Evolve2Stable2} we show $\psi(r,t)$ at different snapshots in time from perturbing by $\pm$2\% away from an exact stable (blue) branch solution, while in Fig.~\ref{Evolve2Unstable2} we show $\psi(r,t)$ at different snapshots in time from perturbing by $\pm$2\% away from an exact unstable (red) branch solution. In the upper plots the perturbation is $-$2\% ($\epsilon=-0.02$) and in the upper plots the perturbation is $+$2\% ($\epsilon=+0.02$).

Fig.~\ref{Evolve2Stable2} shows that no matter how we perturb away from a blue branch solution, whether by increasing $\psi$ or decreasing $\psi$ the solution merely oscillates; indicative of a stable solution. This is to be expected from our effective Hamiltonian, formed within some simple ansatz, plotted earlier in Fig.~\ref{hamiltonianvsradius}. That plot predicted that by perturbing to larger or smaller radius, the solution would just oscillate back towards equilibrium.

On the other hand, Fig.~\ref{Evolve2Unstable2} shows that when we perturb away from a red branch solution, the solution runs away. If we (i) decrease $\psi$ (upper plot) the solution begins to expand in size over time. Conversely, if we (ii) increase $\psi$ (lower plot) the solution begins to shrink in size over time, leading to a catastrophic collapse instability. This again is expected from the effective Hamiltonian of Fig.~\ref{hamiltonianvsradius} as follows: By (i) decreasing $\psi$, but maintaining the shape and hence the radius, Fig.~\ref{solutionsminimizedhamiltonian} shows that we have effectively moved to the left of the red curve; this can be equivalently viewed as being above the red curve, and hence having a radius that is {\em too large} for a given number of particles. Then the effective Hamiltonian of Fig.~\ref{hamiltonianvsradius} indicates that we have moved to the {\em right} of the local maximum and so we should continue to expand over time; and presumably head towards the stable branch at the local minimum. Conversely, by (ii) increasing $\psi$, the same chain of reasoning says that we have moved to the right, or equivalently below, the red curve of Fig.~\ref{solutionsminimizedhamiltonian}, and hence having a radius that is {\em too small} for a given number of particles. This means we have moved to the {\em left} of the local maximum and so we should continue to collapse over time.

\section{Physical Parameters for Axions}\label{Physical}

Having established the stability of a class of solutions, it is useful to convert our results back to ordinary dimensionful variables. In terms of the axion mass $m$ and the PQ scale $f_a$, we can express the maximum number of particles $N_{max}$, the maximum mass $M_{max}$, and the minimum clump size $\rr_{90,min}$, for the stable (blue) branch as follows
\bea
N_{max} \amp = \amp {f_a\over m^2\sqrt{G}}\,\n_{max} \sim 8 \times 10^{59}\,(\tilde{m}^{-2}\tilde{f}_a)\,,\label{Nmaxphysical}\\
M_{max} \amp = \amp N_{max}\,m \,\,\,\,\,\,\,\,\,\,\,\,\,\sim 1.4 \times 10^{19}\,\mbox{kg}\,(\tilde{m}^{-1}\tilde{f}_a)\,,\label{Mmaxphysical}\\
\rr_{90,min} \amp = \amp {a\,(\R_{90}/\R)\over bN_{max}Gm^3} \,\,\,\sim 130\,\mbox{km}\,(\tilde{m}^{-1}\tilde{f}_a^{-1})\label{Rminphysical}\,,
\eea
where $\tilde{f}_a\equiv f_a/(6\times 10^{11}\,\mbox{GeV})$ and $\tilde{m}\equiv m/(10^{-5}\,\mbox{eV})$. These values agree roughly with the results reported by Ref.~\cite{Eby:2014fya}. However, Ref.~\cite{Eby:2014fya} incorrectly identifies the stable and unstable branches, where they claim the spatially smaller clumps are stable and the spatially larger clumps are unstable, when the correct behavior is the exact opposite, as shown here. 

The above maximum number of axions that can be in a clump should be compared to the typical number of axions in inhomogeneous patches in the early universe. Before the QCD phase transition, the axion is effectively massless and its correlation length is of the horizon size due to causality $\xi\sim 1/H_{QCD}\sim M_{Pl}/T_{QCD}^2$ (assuming PQ symmetry breaking is after inflation). Furthermore, the number density of axions at this time is $n=\rho/m\sim (T_{eq}/T_{QCD})\rho_{QCD}/m\sim (T_{eq} T_{QCD}^3)/m$, where $T_{eq}$ is the temperature at matter-radiation equality $\sim 0.1$\,eV. This gives the number of axions within a typical correlation length $N_\xi\sim \xi^3\,n$ as \cite{Guth:2014hsa}
\beq
N_\xi\sim {T_{eq}M_{Pl}^3\over T_{QCD}^3m}\sim 10^{61}\,\tilde{m}^{-1}\,.
\label{Nxi}
\eeq
Note that this value is a factor of $\sim 10$ larger than the maximum number of axions that can be within a clump from Eq.~(\ref{Nmaxphysical}). Since there are fluctuations on a range of scales, one anticipates there are an appreciable number of configurations that can allow a fraction of the axions to eventually re-organize into the above clumps once the gravitational mode-mode interactions becomes faster than Hubble damping.

Note that if we consider ultra-light axions, as may be inspired by string theory, the size and mass of these stable clumps becomes much larger as they scale as $\propto1/m$. At the same time, the rate of formation of the BEC due to gravitational mode-mode interactions have a rate $\Gamma_k\sim 8\pi G m\rho/k^2$ \cite{Sikivie:2009qn,Erken:2011dz}. If we write $k=m\,v$ and treat $v$ as roughly fixed by galactic dynamics (say a typical virial velocity), then the rate naively scales as $\Gamma\propto 1/m$ in the galaxy, which could be quite large leading to rapid formation of such objects.

We end this section by mentioning that the ground state is well described by the weak field gravitational approximation. The stable branch in Fig.~\ref{solutionsminimizedhamiltonian} always maintains values much higher than the Schwarzschild radius $R_S=2\,G\,M$. To show this, we consider the ratio
\beq
{R\over R_S} > {R_{min}\over 2\, G\, M_{max}}={\R_{min}\over 2\,\delta\,\n_{max}}\approx 4\times 10^{12}\,\tilde{f}_a^{-2}\,,
\eeq
which demonstrates that there is no possibility for black hole formation of these low density objects when $f_a\ll M_{Pl}$. Furthermore, these objects will not exhibit strong lensing and may be hard to detect even with micro-lensing. On the other hand, strong field effects can emerge if one were to move away from the traditional QCD axion and investigate extremely high values of $f_a$, approaching the Planck scale \cite{Helfer:2016ljl}.

\section{Dense Branch and Axitons}\label{relativistic}

We mentioned before that the Lagrangian density in the non-relativistic limit does not show  its original invariance under the transformation $\phi \rightarrow \phi + 2\pi f_a$. Then to trust the non-relativistic approximation, we have to be sure that the axion field satisfies the condition   
\beq
\frac{\phiamp}{2\pi f_a} = \frac{\Psi_0}{\pi f_a \sqrt{2m}} = \sqrt{\frac{\delta\,\n}{2 \pi^3 \R^3}} \ll 1\,,
\label{phimax}
\eeq
where $\phiamp$ is the amplitude of the field's oscillations and $\Psi_0=\sqrt{N/(\pi\rr^3)}$ for the exponential ansatz. This condition can be re-expressed as a condition on solutions of Eq.~(\ref{zeroorderinY}) that the radius is bounded from below for the non-relativistic analysis to be self consistent
\beq
\R \gg \Rm(\n) = \left(\frac{\delta\,\n}{2\,\pi^3}\right)^{1/3}\,,
\label{rhomin}
\eeq
where 
\beq
\delta \equiv G f_a^2\,,
\eeq
is the residual parameter in the problem. Note that for the parameters of interest here, $\delta\ll 1$; for example, for $f_a=6\times 10^{11}$\,GeV, it is $\delta\approx 2.5\times 10^{-15}$.

For the blue stable branch this condition is always satisfied. For the red unstable branch this condition is satisfied for most of the branch, except when $\n$ becomes very small. The low $\n$ asymptotic behavior of the red branch in the non-relativistic approximation is given by $\R=3c\,\n/(2a)$, which violates this condition at $\n\lesssim \mathcal{O}(10^{-5})$ for $\delta=2.5\times10^{-15}$. For these small values of $\n$ and $\R$ we need to return to the relativistic theory.

In this corner of phase space, we know that the self-interactions are entirely dominant over gravity, so we can ignore the gravitational corrections. We are interested in periodic clump solutions, which in general can have a tower of harmonics, but will simplify the analysis by allowing only a single frequency $\omega$. This will provide a very rough and only qualitative description of the system. But will be sufficient to convey the qualitative idea for now, and we leave a more precise treatment for future work. A spherically symmetric approximate solution then takes the form
\beq
\phi(r,t)=\Phi(r)\cos(\omega\,t)\,.
\label{relativisticform}
\eeq
We insert this into the Hamiltonian and average over a period of oscillation $T=2\pi/\omega$ as
\beq
\langle H\rangle={1\over T}\int_0^T dt\,H\,.
\eeq
Carrying out this time average in the relativistic Hamiltonian (ignoring gravity, but including the full cosine potential) readily gives
\beq
\langle H\rangle=4\pi\int_0^\infty dr\,r^2\left[{\omega^2\over 4}\Phi^2+{1\over4}\Phi'^2+m^2f_a^2\left[1-J_0(\Phi/f_a)\right]\right]\,,
\eeq
where $J_0$ is the Bessel function of order 0. 

We also need to specify the condition for $\omega$. Consider the equation of motion
\beq
\ddot\phi-\nabla^2\phi+m^2 f_a\sin(\phi/f_a)=0\,.
\eeq
To extract the fundamental frequency, we insert Eq.~(\ref{relativisticform}) into this, multiply by $\cos(\omega\,t)$, time average over a period, and integrate over space to obtain the approximate value
\beq
\omega^2 \approx 2m^2 \left[ \int_0^\infty dr\,r^2\,J_1(\Phi/f_a) \right] \Big{/} \left[ \int_0^\infty dr\, r^2\,(\Phi/f_a) \right]\,,
\label{omegasquared}
\eeq
where we have used the fact that the Laplacian term $\nabla^2\phi$ is a total derivative and so it integrates over space to zero. 

We will continue to use the exponential ansatz for the radial profile for simplicity. We parameterize it as
\beq
\Phi_R(r) = 2\pi\,\varepsilon\,f_a\,e^{-r/R}\,,
\eeq
where the amplitude is specified by $\varepsilon$ that lives in the domain $0<\varepsilon<1$ to ensure that $|\phi|<2\pi\,f_a$ always. Note that these approximations give a frequency that is independent of width $R$ and only depends on its amplitude $\varepsilon$ in Eq.~(\ref{omegasquared}). By evaluating $\omega(\varepsilon)$ we find that the frequency of oscillation is {\em lowered} at finite amplitude from the zero-amplitude $\omega=m$ value.

To evaluate the Hamiltonian with this exponential profile, we use the power series expansion of the Bessel functions
\beq
J_0(x) = \sum_{l=0}^\infty{(-1)^l\over (l!)^2}\left(x\over2\right)^{2l}\,,\,\,\,\,
J_1(x) = \sum_{l=1}^\infty{(-1)^{l-1}\over l!(l-1)!}\left(x\over2\right)^{2l-1}\,,
\eeq
and find that the Hamiltonian is given in terms of generalized hypergeometric functions as follows
\beq
\langle H\rangle = f_a^2\pi^3R\,\varepsilon^2\left(1+m^2R^2\,g(\varepsilon) \right)\,,
\label{hrel}
\eeq
where
\beq
g(\varepsilon)\equiv \, _{3}F_4\!\left({\textstyle 1/2,1/2,1/2 \atop \textstyle 3/2,3/2,3/2,2}\,;-\pi^2\varepsilon^2\right) + \, _{4}F_5\!\left({\textstyle 1,1,1,1 \atop \textstyle 2,2,2,2,2}\,;-\pi^2\varepsilon^2\right)\,.
\eeq

\begin{figure}[t]
\centering
\includegraphics[scale=0.25]{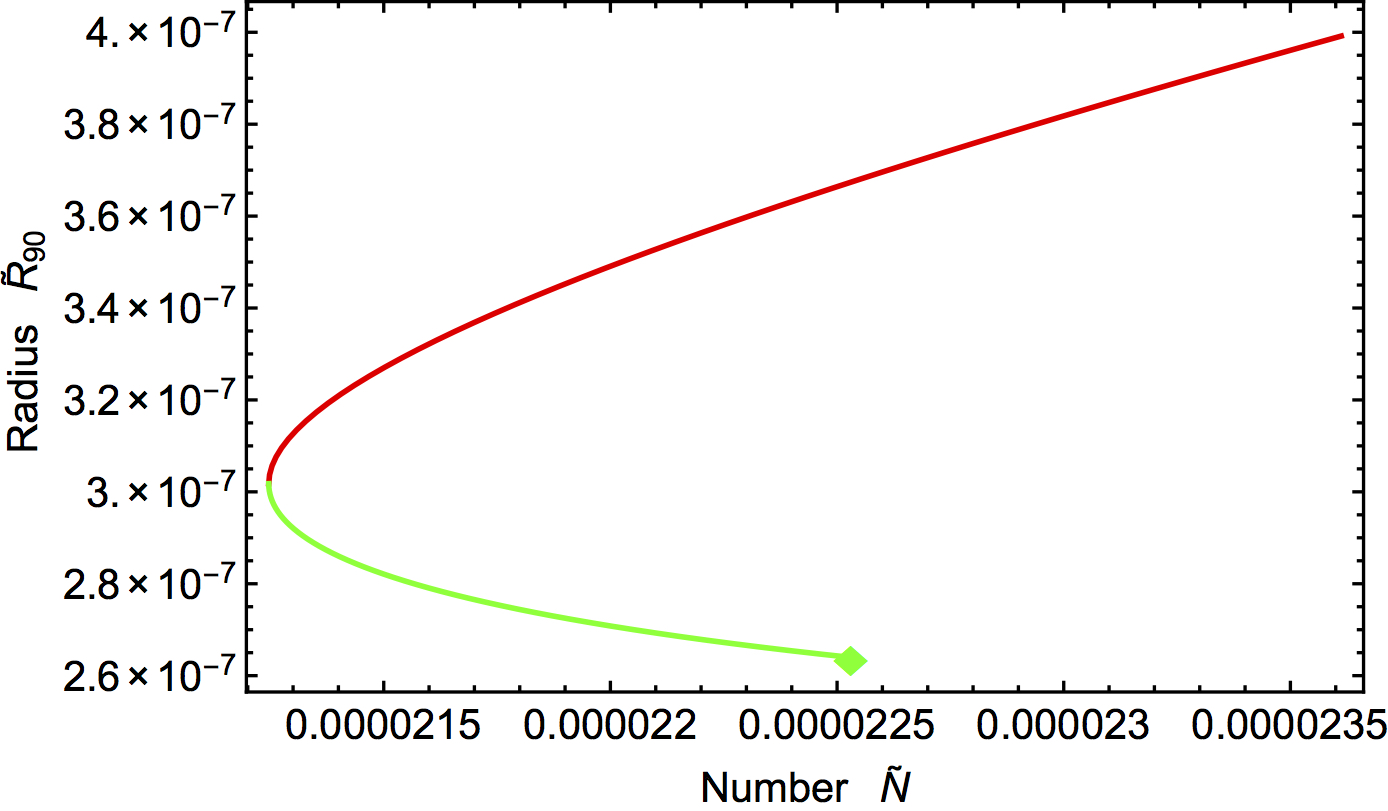}
\caption{Relativistic branch of solutions which break the condition given by Eq.~(\ref{rhomin}). This is computed within the single frequency, exponential profile ansatz, which is only a very rough approximation to the true behavior. The upper red curve are unstable solutions, which, when extrapolated to higher $\n$ and $\R_{90}$, connect to the non-relativistic result of Fig.~\ref{solutionsminimizedhamiltonian}. The lower green curve are quasi-stable solutions, known as ``axitons", which persist only down to the green diamond. We have set $\delta=2.5\times 10^{-15}$ to normalize the number $\n$ and radius $\R_{90}$.}
\label{relativisticbranch}
\end{figure}

We would like to extremize the Hamiltonian as we did earlier at fixed particular number. Strictly speaking the particle number is not conserved in an interacting relativistic theory, but again by time averaging over a period, we have
\beq
\langle N\rangle = \int d^3x\,\omega\langle\phi^2\rangle = 2\pi^3f_a^2 R^3\omega(\varepsilon)\varepsilon^2\,.
\eeq
Using this to eliminate $\varepsilon$ and expressing $\langle H\rangle$ in terms of $\langle N\rangle$ and $\rr$ we can find the extrema numerically. By re-scaling to the same dimensionless variables we used earlier (despite the awkward fact that it now introduces $G$, so the result now depends on our choice of $\delta=G\,f_a^2$) we find the result given in Fig.~\ref{relativisticbranch}. We see that it matches the previous small field result in the upper red branch. Indeed one can check that for large $\rr$ the Hamiltonian in Eq.~(\ref{hrel}) becomes 
\beq
\langle H\rangle=\langle N\rangle\,m+{\langle N\rangle\over 2m R^2}-{\langle N\rangle^2\over 128\pi f_a^2\,R^3}+\ldots\,\,\,\,\,(\mbox{large $\rr$})\,,
\eeq
matching the non-relativistic Hamiltonian derived earlier in Eq.~(\ref{hnonrel}) with an overall shift in energy of $\langle N\rangle\,m$ and without gravity.

While the upper red branch is unstable, as we saw in the non-relativistic limit, the lower green branch is stable. These are truly relativistic solutions with frequencies of oscillations significantly far from $m$; such solutions are known as ``axitons" \cite{Kolb:1993hw}. It turns out such solutions are only quasi-stable as they radiate relativistic axions at an appreciable rate. As they do so, they move on an upper left trajectory in Fig.~\ref{relativisticbranch} until they reach the critical point at which the two branches meet, then implode. Note that these quasi-stable axitons exist in a rather narrow regime of phase space, as their particle number is both bounded above (by the requirement that the field $\phiamp<2\pi f_a$, indicated by the green diamond) and bounded below (by the requirement that it is stable, indicated by the point at which it meets the red curve). Both the upper and lower bounds are of the same order, given roughly by $N\sim f_a^2/m^2$, with $R\sim 1/m$. On the other hand, the stable gravitational solutions found in the previous subsection have only an upper bound on $N$ (and only a lower bound on $\rr$) and therefore occupy a much larger portion of phase space.

We would like to contrast these results to the work of Refs.~\cite{Braaten:2015eeu,Eby:2016cnq}, where the authors claim there exist a stable dense branch that extends to arbitrarily large particle number $\n$, all within the non-relativistic framework. However, such a result is erroneous, as it is an artifact of improper usage of the non-relativistic theory in a regime of arbitrarily large amplitude, which breaks the condition of Eq.~(\ref{phimax}) and does not enforce periodicity of the field. Instead this dense branch requires the above relativistic treatment and exhibits a final endpoint of the green curve as indicated here.

\section{Repulsive Self-Interactions }\label{Repulsive}

Another class of behavior occurs if we move away from the QCD axion, which is organized by an {\em attractive} $-\lambda\,\phi^4$ interaction from expanding the cosine potential, to a generic light scalar dark matter candidate, that may be described by a {\em repulsive} $+\lamr\,\phi^4$ interaction (with $\lamr>0$). A simple (renormalizable) potential is
\beq
V(\phi)={1\over2}m^2\phi^2+{\lamr\over4!}\phi^4\,.
\eeq
So long as the particle is sufficiently light such that the number density, and hence occupancy number, is large to comprise the dark matter, we can again study this within classical field theory. 

\begin{figure}[t]
\centering
\includegraphics[scale=0.23]{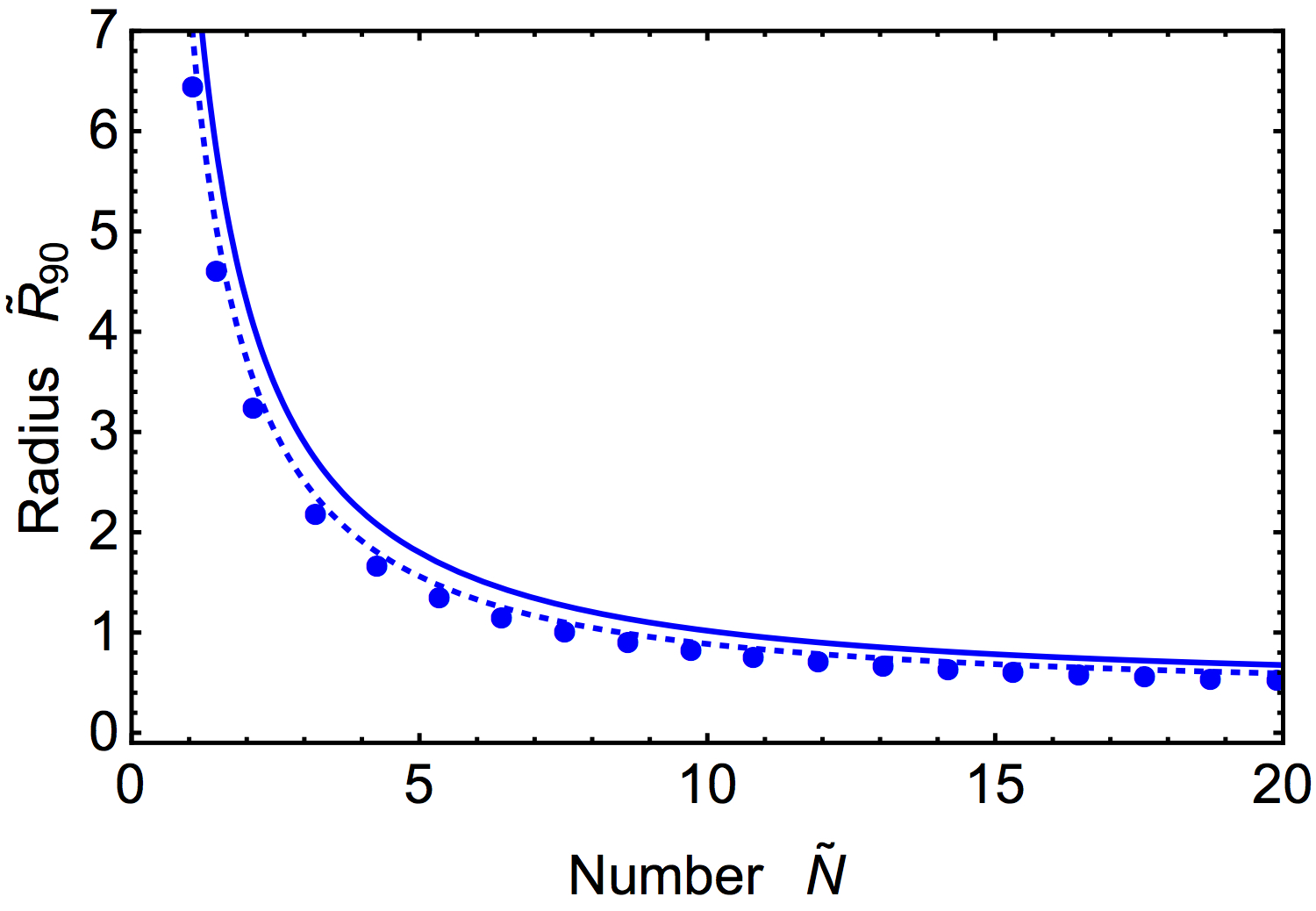}
\caption{Single (stable) solution branch for repulsive self-interactions for generic scalar dark matter. The system is treated in the non-relativistic regime. We have plotted the dimensionless radius $\R_{90}$ (defined as the radius that encloses 90\% of the mass) versus dimensionless particle number $\n$. Solid curve is the exponential approximation, dotted curve is the sech approximation, and the individual dots are the exact numerical values.}
\label{FigureRadiusNumberRepulsive}
\end{figure}

In the non-relativistic regime, this leads to exactly the same set of equations as we described earlier, with only the sign of the quartic self-interaction changed
\beq
V_{nr}(\psi,\psi^*) = \lamr{\psi^{*2}\psi^2\over 16\,m^2}\,.
\eeq
We again pass to the dimensionless variables of Eqs.~(\ref{rescR},\,\ref{rescN},\,\ref{rescH}), with the replacement $f_a\to m/\sqrt{\lamr}$. For any localized clump ansatz of a single length scale $\R$, we have an obvious modification in the Hamiltonian from Eq.~(\ref{hamiltonianquadraticgeneral}) to
\beq
\E(\R) \approx a\frac{\n}{\R^2} - b\frac{\n^2}{\R} + c\frac{\n^2}{\R^3 }\,,
\label{hamiltonianquadraticgeneralrepulsive}
\eeq
where the sign of the final (self-interaction) term is flipped. Unlike the previous case of attractive interactions, where there were two branches of extrema, here there is only {\em one} branch of extrema, which is stable, and given by
\beq
\R = {a+\sqrt{a^2+3bc\n^2}\over b\n}\,,
\eeq
(the other branch would correspond to an unphysical negative radius). For the exponential and sech ansatzes described in the earlier sections, we plot this result in Fig.~\ref{FigureRadiusNumberRepulsive}, along with the exact numerical result given by the individual dots. Evidently, this branch extends to arbitrarily large particle number $N$, unlike the previous attractive case where there was an $N_{max}$. In the $N\to\infty$ limit, these simple analytical ansatzes predict that the radius of the clump becomes a fixed value
\beq
\R \to \sqrt{3c\over b}\,\,\,\,\,(\mbox{large $N$})\,.
\eeq
Numerical studies indicate while the radius is almost constant at large $N$, there may still be a slow decrease in $R$ as we increase $N$, in a fashion that is not fully captured by these simple ansatzes.

\section{Summary and Outlook}\label{Conclusions}

In this paper we have mapped out the basic solutions of the axion-gravity-self-interacting system. We established two branches of solutions that exist primarily in the non-relativistic regime, whose relationship between clump size and particle number is given in Fig.~\ref{solutionsminimizedhamiltonian}, with the upper branch stable and the lower one unstable to collapse. We also showed that the lower (unstable) branch connects to a relativistic branch given in Fig.~\ref{relativisticbranch}, which is quasi-stable since it can steadily emit relativistic axions. This quasi-stable ``axiton" branch is to be contrasted with the claims of a dense branch of arbitrary particle number that has appeared in the literature recently \cite{Braaten:2015eeu,Eby:2016cnq}, where the axion's field range was erroneously allowed to be arbitrarily large within the non-relativistic treatment. In any case, the stable (blue) branch of Fig.~\ref{solutionsminimizedhamiltonian} is perhaps of most importance; it is primarily organized by gravity and could comprise an important component of axion dark matter in the galaxy. (In the literature, such clumps are sometimes called "Bose stars" or "oscillatons" and can organize into ``miniclusters"). 

We found that the typical number of axions in a clump is comparable to the typical number of axions in one coherence length in the early universe in the scenario in which the PQ phase transition occurs after inflation. These clumps may potentially also form in the scenario in which the PQ phase transition occurs before inflation, even though the axion field then begins with much less power on small scales, but this can grow appreciably over cosmic time. 

We have made use of a simple ansatz for the shape of the clump, wherein its shape is controlled by a single length scale $R$. We primarily exploited the exponential ansatz for simplicity, though we compared to a sech ansatz, finding similar results. We then proceeded to compute the shape of these clumps numerically and computed their exact time evolution; establishing conclusively that the stable branch is well behaved when perturbed about equilibrium, while the unstable branch either collapses or expands depending on the form of the perturbation. A future direction is to analytically determine the exact shape of these clumps using a systematic expansion. (Some work to establish a series expansion in the pure gravity case is in Ref.~\cite{Kling:2017mif}). 

It is important to compute the actual abundance and distribution of these clumps in the universe and within the galaxy. To do so properly, it would be important to perform numerical simulations, including FRW expansion. Although our estimates suggest that such clumps are stable and are built out of an axion number that seems to appear naturally, it would be important to see this play out in simulations. It would also be interesting to see if the field primarily organizes into ground state clumps, as suggested by BEC arguments \cite{Guth:2014hsa}, or if higher angular momentum eigenstates that break the spherical symmetry assumption survive over time, which will be addressed in a further study \cite{Forthcoming}.

These findings could have implications for direct detection strategies: if the axion organizes primarily into bound clumps, then it may reduce the probability of dark matter axions passing through earth based detectors. This may invite alternative search strategies, such as femtolensing/picolensing \cite{Kolb:1995bu}, or to explore possible resonant coupling to photons from these clumps, which will also be addressed in a further study \cite{Forthcoming}.

We also examined more generic scalar dark matter, allowing for repulsive self-interactions, which has only a stable clump solution branch that extends to  arbitrarily large particle number and is rather compact. This may have interesting astrophysical consequences.

\section*{Acknowledgments}

For discussions on these topics, we would like to thank Alan Guth, Jose Blanco-Pillado, Mohammad Hossein Namjoo, Chanda Prescod-Weinstein. MPH is supported by National Science Foundation grant PHY-1720332.

\appendix

\section{Appendix: Instability from Homogeneous Field}\label{AppendixA}

Let us consider the case in which the scalar field is initially homogeneous. This can arise from having the field arise from a phase transition after inflation. We would like to analytically compute the behavior of perturbations. For completeness, we will perform a fully relativistic analysis here.

Recall the full relativistic Lagrangian density, of Eq.~(\ref{axionlagrangiandensity}). We will work in the weak field limit, and write the Newtonian metric as $g_{00}= 1+2\phi_N$, $g_{ij}=-(1-2\phi_N)\delta_{ij}$, $g_{0i}=0$. By varying the action and working to linear order in the Newtonian potential $\phi_N$, the equations of motion are
\bea
&&\frac{d^2\phi}{dt^2} - \nabla^2\phi -2\phi_N \frac{d^2\phi}{dt^2}-2\phi_N \nabla^2\phi -4 \frac{d\phi_N}{dt} \frac{d\phi}{dt}+\frac{dV(\phi)}{d\phi}=0\,,\label{axionequationofmotion}\\
&&\nabla^2\phi_N = 4\pi G \left[ \frac{1}{2}\left(\frac{d{\phi}}{dt}\right)^2 + V(\phi) \right]\,.\label{newtonianpotential}
\eea
where we have dropped corrections on the right hand side of Eq.~(\ref{newtonianpotential}) as they are suppressed in the weak field regime.

\subsection{Background}

Let us denote the background field $\phiamp$, which we take to be a function of time only. The equation of motion for $\phiamp$ is simply
\beq
\frac{d^2\phiamp}{dt^2} + \frac{dV(\phiamp)}{d\phiamp} = 0\,,
\label{axioneqm}
\eeq
where the potential $V(\phi)$ is given by Eq.~(\ref{potentialexpansion}). The mass term in $V(\phi)$ dominates the oscillatory behavior of the background field, $\phiamp(t)$,  leading to almost harmonic motion.  The self-interacting terms will give rise to anharmonic behavior and eventually could drive resonance in perturbations. 

To make progress, we take small field values and expand the potential $V(\phi)$ as follows
\beq
V(\phi) = \frac{1}{2}m^2\phi^2 - \frac{\lambda}{4!}\phi^4 + \frac{g}{6!}\phi^6 + \ldots\,.
\label{potentialexpansion}
\eeq
Here the dots indicate terms of $ \mathcal{O}(\phi^8)$. For convenience we will often write $g \equiv g'\lambda^2/m^2$. For axions, we have $\lambda \equiv (\Lambda/f_a)^4>0$ and $g'=1$. For generic scalar dark matter, we can consider $\lambda=-\lamr<0$ and/or consider $g' \not= 1$.

The background field  can be considered in a small amplitude expansion as
\beq
\phiamp = \varepsilon \phi_1 + \varepsilon^3 \phi_3 + \varepsilon^5 \phi_5\,...\,, 
\label{expansionbackground}
\eeq
where $\varepsilon$ is a small dimensionless constant. As we will see, replacing this expansion into the equation of motion and naively matching powers of $\varepsilon$ would lead to an undesired secular behavior (when the driving terms match the natural frequency defined by the harmonic terms). 

To avoid this problem, we introduce a new time variable $\tau \equiv \sqrt{1 - \varepsilon^2}\,t$ to describe the shifted frequency, since the attractive self-interaction diminishes the fundamental oscillation frequency. (Note that we are describing the effects for $\lambda>0$, but one can easily send $\lambda\to-\lamr$ and $\varepsilon\to i\varepsilon$ to obtain the results for repulsive scalar dark matter.) Then Eq.~(\ref{axioneqm}) becomes  
\beq
\phiamp'' \left( 1- \varepsilon^2 \right) + m^2\phiamp - \frac{\lambda}{6}\phiamp^3 + \frac{g}{120}\phiamp^5 +\ldots= 0\,,
\label{equationofmotionintau}
\eeq
where the prime means derivatives with respect to $\tau$. Replacing the expansion of Eq.~(\ref{expansionbackground}) into this equation and matching terms until first order in $\varepsilon$, we obtain
\beq
\phi_1'' + m^2 \phi_1 = 0\,,
\eeq
whose solution (up to a phase) is given by
\beq
\phi_1 = \phi_{1a} \cos({m\tau})\,.
\label{phi1}
\eeq
Here the value for the amplitude $\phi_{1a}$ has to be determined. Replacing again Eq.~(\ref{expansionbackground}) into Eq.~(\ref{equationofmotionintau}), but now working to $\mathcal{O}(\varepsilon^3)$, we obtain 
\beq
\phi_3'' + m^2\phi_{1a}\cos({m\tau}) + m^2\phi_3 - \frac{\lambda}{24}\phi^3_{1a}\left[ \cos(3m\tau) + 3\cos(m\tau) \right]=0\,. 
\eeq
To avoid secular behavior, we have to eliminate the factor proportional to $\cos(m\tau)$ which selects a unique value for the amplitude $\phi_{1a}$. This procedure, explained in detailed in Refs.~\cite{Hertzberg:2014iza,Hertzberg:2014jza}, can be extended as many orders in $\varepsilon$ as we want. Until $\mathcal{O}(\varepsilon^3)$, the background solution is given by
\beq
\phiamp = \varepsilon \phi_{1a} \cos(m\tau) + \varepsilon^3 \left[ \phi_{3a}\cos(m\tau) - \phi_{3b}\cos(3m\tau)  \right]\,,
\label{backgroundinepsilonpowers}
\eeq
where 
\beq
\phi_{1a} = \sqrt{8m^2\over\lambda},\,\,\, \phi_{3a}={\phi_{1a}^3\over384}{\lambda\over m^2}(1 + 8g'),\,\,\, \phi_{3b}={\phi_{1a}\over24}\,.
\eeq

\subsection{Perturbations}

Now that we have determined the background to the order of interest, we expand the axion field around this classical background as
\beq
\phi({\bf{x}},t) = \phiamp(t) + \dphi({\bf{x}},t)\,.
\label{perturbationfield}
\eeq
Here we will treat $\dphi({\bf{x}},t)$ as a classical perturbation (though it could be generated by quantum fluctuations). We work to first order in $\dphi({\bf x},t)$ and $\phi_N({\bf x},t)$. Their equations of motion can be easily diagonalized by Fourier transforming to ${\bf k}$-space variables $\dphi_{\bf k}(t)$ and $\phi_{N{\bf k}}(t)$, respectively. We then obtain the following pair of coupled equations for $\dphi_{\bf k}$ and $\phi_{N{\bf k}}$
\bea
&&\frac{d^2}{dt^2}\dphi_{\bf k}+ k^2\delta{\phi_{\bf{k}}} - 2\frac{d^2 \phi_0}{dt^2}\phi_{N \bf{k}} - 4 \frac{d\phi_0}{dt}{d\over dt}\phi_{N{\bf k}}+\frac{d^2V(\phi)}{d\phi^2}\dphi_{\bf k} =0\,,\label{perturbationeqmkspace}\\
&&\phi_{N {\bf{k}}} = -\frac{4\pi G}{k^2}\left( \frac{d\phi_0}{dt}\frac{d}{dt}\dphi_{\bf k}+\frac{dV(\phiamp)}{d\phiamp}\dphi_{\bf k} \right) \,.
\eea
We can analyze the late time behavior of the system using Floquet theory. Floquet exponents rule any possible exponential growth of perturbations around a background which is oscillating periodically. This is a very reasonable approximation in the limit when the oscillation period (which is of the order of $2\pi/m$)  is short compared to the Hubble time. The Floquet theory still involves only numerical solutions; here we will provide analytical results at small amplitudes.

By eliminating $\phi_{N{\bf k}}$, the equation of motion for the perturbation $\dphi_{\bf k}$ can be expressed in the following form
\beq
h_1(\tau)\,\delta\phi''_{\bf{k}} + h_2(\tau)\,\delta\phi'_{\bf{k}} + h_3(\tau)\,\delta\phi_{\bf{k}} =0\,,
\label{Hillequation1}
\eeq
where $h_{1,2,3}$ are all periodic functions of the re-scaled time variable $\tau$.

\subsection{First Instability Band}

Working to $\mathcal{O}(\varepsilon^2)$, which involves only needing the background solution to $\mathcal{O}(\varepsilon)$, we obtain the following expression for each of these 3 coefficients
\bea
&&h_1(\tau) = A + B \cos({2m\tau}),\\
&&h_2(\tau) = C\sin({2m\tau}),\\
&&h_3(\tau) = D + E \cos({2m\tau})\,.
\eea
Here $A, B, C, D$ and $E$ are given by
\bea
A \amp = \amp 1 + \frac{8\pi G\phi_{1a}^2m^2\varepsilon^2}{k^2}\,,\label{A}\\
B \amp = \amp -\frac{8\pi G\phi_{1a}^2m^2\varepsilon^2}{k^2}\,,\label{B}\\
C \amp = \amp \frac{4\pi G\phi_{1a}^2m^3\varepsilon^2}{k^2}\,,\label{C}\\
D \amp = \amp k^2 + m^2 - m^2\varepsilon^2 +\frac{4\pi G\phi_{1a}^2m^4\varepsilon^2}{k^2}\,,\label{D}\\
E \amp = \amp -2m^2\varepsilon^2-\frac{12\pi G\phi_{1a}^2m^4\varepsilon^2}{k^2}\,.\label{E}
\eea
We have kept terms until $\mathcal{O}(\varepsilon^2)$ taking into account that $k\sim \mathcal{O}(\varepsilon)$ and $G\sim \mathcal{O}(\varepsilon)$. 

To solve for the time evolution, lets express the solutions for perturbations as a harmonic expansion over integer multiplies of the fundamental frequency $m$ as follows
\beq
\delta \phi_{\bf{k}} = \sum_{\omega = -\infty}^{+\infty} e^{i\omega\tau}f_{\omega}(\tau)\,.
\label{harmonicexpansionmathieequation}
\eeq
Assuming that functions $f_{\omega}(\tau)$ are slowly varying, we drop second derivative terms $f_\omega''$, to obtain the following infinite system of coupled ordinary differential equations 
\bea
\amp  2iA\omega f_{\omega}'(\tau)  +  i\left[ B(\omega + 2m)+\frac{c}{2}\right]f_{\omega + 2m}'(\tau) +i\left[B(\omega - 2m) - \frac{c}{2}\right]f_{\omega - 2m}'(\tau)+  (-\omega^2 A+D)f_{\omega}(\tau)  \nonumber \\
\amp -\left[\frac{B(\omega + 2m)^2+C(\omega+2m)-E}{2}\right]f_{\omega + 2m}(\tau) - 
    \left[\frac{B(\omega - 2m)^2-C(\omega-2m)-E}{2}\right]f_{\omega - 2m}(\tau)=0\,.\,\,\,\,\,\,
\label{ODE}
\eea
Note that only odd (even) harmonics couple with odd (even) harmonics.  

The fundamental frequencies, $\omega = \pm m$, give us information about the first instability band. To leading order, we can drop higher harmonics in Eq.~(\ref{ODE}), $f_{\pm 3m}(\tau)$, to obtain the following pair of coupled ordinary differential equations for $f_{\pm m}(\tau)$
\beq
\left[ \begin{array}{c} f'_m(\tau) \\ f'_{-m}(\tau) \end{array} \right] =  \mathcal{M}_1  \left[ \begin{array}{c} f_m(\tau) \\ f_{-m}(\tau) \end{array} \right]\,,
\label{matrixequation}
\eeq
where the matrix $\mathcal{M}_1$ is given by
\beq
\mathcal{M}_1 \equiv \frac{i}{(2mA)^2-\left(mB+\frac{c}{2}\right)^2}\begin{bmatrix} -2mA & mB+\frac{c}{2} \\ -\left(mB+\frac{c}{2}\right) & 2mA \end{bmatrix} \times \begin{bmatrix} m^2A-D & \frac{(m^2B+cm-E)}{2} \\ \frac{(m^2B+cm-E)}{2} & m^2A-D \end{bmatrix}\,.
\eeq

The general solutions for $f$ take on the form
\beq
f_{\pm}=c_1\,e^{\mu_k \tau}+c_2\,e^{-\mu_k \tau}
\eeq
where the exponents, $\pm \mu_k$, are the eigenvalues of the above matrix. A non-zero real part of $\mu_k$ leads to an exponential growth of perturbations. By contrast, a purely imaginary Floquet exponent produces an oscillatory behavior of perturbations (stable time evolution). The (positive) eigenvalue of the matrix in Eq.~(\ref{matrixequation}) are 
\beq
\mu_k = \frac{\sqrt{Cm-E-2D+2Am^2+Bm^2}\,\sqrt{2D-E+Cm-2Am^2+Bm^2}}{\sqrt{4Am-C-2Bm}\,\sqrt{4Am+C+2Bm}}\,.
\eeq
Defining $\phi_a \equiv \varepsilon\phi_{1a}$ as the physical amplitude and replacing values for $A, B, C, D$ and $E$ from Eqs.~(\ref{A} -- \ref{E}) into this expression for $\mu_k$, we obtain at this order
\bea
\mu_k \amp = \amp \frac{k}{2m}\sqrt{\frac{\phi_a^2\lambda}{4}-k^2+\frac{8\pi G \phi_a^2 m^4}{k^2}}\,.
\label{Floquetexponent}
\eea
So there is an instability band with edges given by values for $k$ at which the Floquet exponent becomes zero. The left ($k_{l,edge} $) and right ($k_{l,edge}$) hand edges are calculated to be
\bea
k_{l,edge} \amp = \amp 0\,,\label{kleft}\\
k_{r,edge} \amp = \amp {\phi_a\sqrt{\lambda+\sqrt{512\pi Gm^4+\lambda^2}}\over2\sqrt{2}}\label{kright}\,.
\eea
As a result, we expect a dominant thick band that extends to $k=0$ with an exponent that is linear in $k$ in the long wavelength regime. We note that the shut-off of the instability at $k_{r,edge}$ defines a type of Jeans wavenumber.

Note that the results in Eqs.~(\ref{Floquetexponent},\,\ref{kleft},\,\ref{kright}) are true for both $\lambda>0$ and $\lambda=-\lamr<0$. In the axion case of $\lambda > 0$ we can go further and see there is a non-zero wavenumber $k^* = (\phi_a \sqrt{\lambda})/(2\sqrt{2})$ that maximizes the exponential growth of perturbations as 
\beq
\mu^* = \frac{\phi_a^2\lambda}{16m}\sqrt{1+\frac{512\pi G m^4}{\lambda^2 \phi_a^2} }\,.
\eeq
While if $\lambda=-\lamr<0$ the growth is maximal as $k\to0$ with value $\mu^*=\sqrt{2\pi G}\,\phi_a\,m$.

\subsection{Second Instability Band}

To study the second instability band, we repeat the above procedure but work to $\mathcal{O}(\varepsilon^4)$, which involves needing the background solution to $\mathcal{O}(\varepsilon^3)$. We find $h_{1,2,3}$ in Eq.~(\ref{Hillequation1}) are 
\bea
h_1(\tau) \amp = \amp A + B\cos(2m\tau)+C\cos(4m\tau)\,,\\
h_2(\tau) \amp = \amp D \sin(2m\tau) + E \sin(4m\tau)\,,\\
h_3(\tau) \amp = \amp F + J \cos(2 m\tau) + H \cos(4 m\tau)\,,
\eea
where $A,B,C,D,F,J$, and $H$ are given by
\bea
A \amp = \amp 1 + \frac{8\pi G \phi_{1a}^2 m^2 }{k^2}\varepsilon^2-\left(\frac{8\pi G \phi_{1a}^2 m^2}{k^2}-\frac{g\pi G \phi_{1a}^4 m^2}{3k^2\lambda}-\frac{\pi G \phi_{1a}^4 \lambda}{24k^2}\right)\varepsilon^4\,,\label{A2}\\
B \amp = \amp -\frac{8\pi G \phi_{1a}^2 m^2}{k^2}\varepsilon^2 + \left( \frac{6\pi G \phi_{1a}^2m^2}{k^2} - \frac{g\pi G \phi_{1a}^4 m^2}{3\lambda k^2} - \frac{\pi G \phi_{1a}^4\lambda}{24k^2}  \right)\varepsilon^4\,,\label{B2}\\
C \amp = \amp \frac{2\pi G \phi_{1a}^2 m^2 }{k^2}\varepsilon^4 \,,\label{C2}\\
D \amp = \amp \frac{4\pi G \phi_{1a}^2 m^3 }{k^2}\varepsilon^2 - \left( \frac{3\pi G \phi_{1a}^2m^3}{k^2} - \frac{g\pi G \phi_{1a}^4 m^3}{6\lambda k^2} - \frac{\pi G \phi_{1a}^4 \lambda m}{48k^2}  \right)\varepsilon^4 \,,\label{D2}\\
E \amp = \amp -\frac{2\pi G \phi_{1a}^2 m^3 }{k^2}\varepsilon^4 \,,\label{E2}\\
F  \amp = \amp  k^2 + m^2 +  \left( k^2 + m^2 + \frac{4\pi G \phi_{1a}^2 m^4}{k^2} - \frac{\phi_{1a}^2\lambda}{4} \right)\varepsilon^2 + \nonumber \\
 \amp \amp  + \left( k^2 + m^2 - \frac{23 \pi G \phi_{1a}^4\lambda m^2}{48k^2} + \frac{g\pi G \phi_{1a}^4 m^4}{6\lambda k^2}  + \frac{g\phi_{1a}^4}{192} - \frac{\phi_{1a}^2\lambda}{4} - \frac{\phi_{1a}^4\lambda^2}{768m^2} \right)\varepsilon^4 \,,\label{F2}\\
J  \amp = \amp \left( -\frac{12 \pi G \phi_{1a}^2 m^4}{k^2} - \frac{\phi_{1a}^2\lambda}{4}\right)\varepsilon^2 + \nonumber\\
\amp \amp \left( -\frac{\pi G \phi_{1a}^2 m^4}{3k^2} + \frac{29\pi G \phi_{1a}^4 \lambda m^2}{48k^2} - \frac{g\pi G \phi_{1a}^4 m^4}{2\lambda k^2}   + \frac{g\phi_{1a}^4}{96} - \frac{11\phi_{1a}^2\lambda}{48} -\frac{\phi_{1a}^4\lambda^2}{768m^2}  \right)\varepsilon^4\,,\label{J2} \\
H  \amp = \amp \left( \frac{11\pi G \phi_{1a}^2m^4}{3k^2} + \frac{7\pi G \phi_{1a}^4\lambda m^2}{6k^2} +\frac{\lambda\phi_{1a}^2}{48}+ \frac{g\phi_{1a}^4}{192} \right)\varepsilon^4 \,.\label{H2} 
\eea
We replace the harmonic expansion of Eq.~(\ref{harmonicexpansionmathieequation}) into the equation of motion for perturbations, dropping second derivatives as before, to obtain
\bea
\amp 2iA\omega f_{\omega}'+i\left[ \frac{D}{2} + B(\omega + 2m) \right]f_{\omega + 2m}' + i\left[\frac{E}{2}+C(\omega + 4m) \right]f_{\omega + 4m}'-i\left[\frac{D}{2}-B(\omega -2m)  \right]f_{\omega-2m}' \nonumber\\
\amp - i\left[\frac{E}{2}-C(\omega-4m) \right]f_{\omega-4m}'+\left[\frac{J + D(\omega-2m)-B(\omega-2m)^2}{2} \right]f_{\omega-2m}+\left[ \frac{J - D(\omega + 2m)-B(\omega + 2m)^2}{2} \right]f_{\omega+2m} \nonumber\\
\amp +\left[ \frac{H-E(\omega + 4m)-C(\omega + 4m)^2}{2}\right] f_{\omega + 4m}+(F-A\omega^2)f_{\omega}+\left[ \frac{H+E(\omega-4m)-C(\omega-4m)^2}{2} \right]f_{\omega-4m} = 0\,.
\label{ODE2}
\eea
Now the frequencies $\omega = -2m, 0, +2m$ give us information about the second instability band. To leading order, we can drop higher harmonics in Eq.~(\ref{ODE2}), $f_{\pm 4m}(\tau)\,, f_{\pm 6m}(\tau)$. We then solve for $f_0(\tau)$ in terms of $f_{\pm 2m}(\tau)$ to obtain the following coupled pair of ordinary differential equations for $f_{\pm 2m}(\tau)$
\beq
\left[ \begin{array}{c} f'_{2m}(\tau) \\ f'_{-2m}(\tau) \end{array} \right] = \mathcal{M}_2 \left[ \begin{array}{c} f_{2m}(\tau) \\ f_{-2m}(\tau) \end{array} \right]\,.
\label{matrixequation2}
\eeq
where the matrix $\mathcal{M}_2$ is given by
\beq
\mathcal{M}_2 \equiv \frac{i}{(X)^2-\left(Y\right)^2}\begin{bmatrix} -X & Y \\ -Y & X \end{bmatrix} \times \begin{bmatrix} W & Z \\ Z & W \end{bmatrix}\,,
\eeq
with
\bea
X \amp = \amp  4Am -  \frac{DBm^2}{F} - \frac{D^2m}{2F} - \frac{JBm}{F} \,,\\
Y \amp = \amp  2mC + \frac{DBm^2}{F}+\frac{D^2m}{2F} - \frac{DJ}{2F} - \frac{JBm}{F} + \frac{E}{2} \,,\\
W \amp = \amp 4Am^2 - F - \frac{JBm^2}{F}-\frac{JDm}{2F}+\frac{J^2}{4F}\,,\\
Z \amp = \amp 2Cm^2 + E m - \frac{JBm^2}{F} - \frac{JDm}{2F}+\frac{J^2}{4F}-\frac{H}{2}\,.
\eea
The evolution of the system is governed by the eigenvalues of this matrix $\pm\mu_k$. We find them to be
\beq
\mu_k =\frac{ \sqrt{\mu_{k,1}}\, \sqrt{\mu_{k,2}}\, \sqrt{\mu_{k,3}}}{  \sqrt{\mu_{k,4}}\,\sqrt{\mu_{k,5}}} \,,
\label{Floquetexponent2}
\eeq
where
\bea
\mu_{k,1} \amp = \amp F \,,\\
\mu_{k,2} \amp = \amp  2F-H+2Em-8Am^2+4Cm^2 \,,\\
\mu_{k,3}  \amp = \amp  -2F^2+J^2-FH+2EFm-2DJm+8AFm^2+4CFm^2-4BJm^2 \,,\\
 \mu_{k,4} \amp = \amp  EF-DJ+8AFm+4CFm-4BJm \,,\\
 \mu_{k,5} \amp = \amp  -EF+DJ-2D^2m+8AFm-4CFm-4BDm^2 \,.
\eea

The exponent becomes zero when $\sqrt{\mu_{k,1}}\,\sqrt{\mu_{k,2}}\, \sqrt{\mu_{k,3}} = 0$. Replacing expressions for $A,B,C,D,$ $E,F,J$ and $H$ from Eqs.~(\ref{A2} -- \ref{E2}), and working to $\mathcal{O}(\varepsilon^4)$, we find that the left and right hand edge of the instability band are given by
\bea
k_{l,edge} = \sqrt{3}m - \frac{\sqrt{3}}{24m}(\lambda \phi_a^2) - \left(\frac{1+g'}{2}  \right)\frac{\sqrt{3}}{1152m^3}(\lambda \phi_a^2)^2 + \frac{14\pi}{3\sqrt{3}}( Gm\phi_a^2)\,,\label{kedgeleft}\\
k_{r,edge} = \sqrt{3}m - \frac{\sqrt{3}}{24m}(\lambda \phi_a^2) + \left(\frac{1-3g'}{2}  \right)\frac{\sqrt{3}}{1152m^3}(\lambda \phi_a^2)^2 + \frac{14\pi}{3\sqrt{3}}( Gm\phi_a^2)\,,\label{kedgeright}
\eea
where we have used the physical amplitude, $\phi_a$, and $g = g' (\lambda^2/m^2)$. Here the definition of left and
right hand edge is arbitrary because it depends on the sign of $1-g'$. In the relativistic theory, since the homogeneous background is a dense condensate of bosons, quartic interactions can lead to annihilations ($4\phi \rightarrow 2\phi$). For kinematics, in the small amplitude limit, we expect outgoing particles with a wavenumber given by $\sqrt{3}m$, the value at which the second instability starts at the limit when $\phi_a \rightarrow 0$, Eqs.~(\ref{kedgeleft},~\ref{kedgeright}). 

Now the width of the second instability band is 
\bea
\Delta k = |k_{r,edge}-k_{l,edge}| = |1 - g'|\frac{\sqrt{3}}{1152m^3}(\lambda \phi_a^2)^2\,.
\eea
For axions, since $g'=1$, we have $\Delta k = 0$ and there is no second instability band. By contrast, for generic scalar dark matter with $g' \neq 1$ there can be a second instability band. If we parameterize moving through the band as $k=k_{l,edge}+\delta k$ (with $|\delta k|<\Delta k$) there is a nonzero real value for $\mu$ to induce exponential growth given by
\beq
\mu  = \frac{\sqrt{\delta k\left[(1-g')m-6\sqrt{3}~\delta k  \right]  }}{128 \sqrt{2}~3^{1/4}m^4}\left( \lambda \phi_a^2 \right)^2\,.
\eeq

\end{document}